\documentclass[aps,showkeys,preprintnumbers,amsmath,epsf,amssymb,epsfig,nofootinbib]{revtex4}
\usepackage{bm}
\usepackage{graphicx}
\newcommand {\eps}{\epsilon}

\newcommand {\si}{\sigma}
\newcommand {\ga}{\gamma}
\newcommand {\Ga}{\Gamma}
\newcommand {\de}{\delta}

\newcommand {\la}{\lambda}

\newcommand {\pa}{\partial}
\newcommand {\na}{\nabla}

\newcommand {\fr}{\frac}

\newcommand {\ca}{{\cal A}}

\newcommand {\cf}{{\cal F}}
\newcommand {\cg}{{\cal G}}
\newcommand {\ch}{{\cal H}}

\newcommand {\co}{{\cal O}}

\newcommand {\cx}{{\cal X}}

\newcommand {\wh}{\hat}
\newcommand {\pe}{\perp}
\newcommand {\lan}{\langle}
\newcommand {\ran}{\rangle}
\newcommand {\mbf}{\mathbf}
\newcommand {\bk}{\mathbf{k}}
\newcommand {\bR}{\mathbf{R}}
\newcommand {\mZ}{\mathfrak{J}}

\newcommand {\beg}{\begin{equation}}

\newcommand {\en}{\end{equation}}
\newcommand {\bega}{\begin{eqnarray}}

\newcommand {\ena}{\end{eqnarray}}

\begin{document}
\title{Seminal magnetic fields from Inflato-electromagnetic Inflation}
\author{$^{1,2}$ Federico Agust\'{\i}n Membiela\footnote{E-mail address:
membiela@mdp.edu.ar} and $^{1,2}$ Mauricio Bellini
\footnote{E-mail address: mbellini@mdp.edu.ar} }
\address{$^1$ Departamento de F\'isica, Facultad de Ciencias Exactas y
Naturales, Universidad Nacional de Mar del Plata, Funes 3350, C.P.
7600, Mar del Plata, Argentina.\\
$^2$ Instituto de Investigaciones F\'{\i}sicas de Mar del Plata (IFIMAR), \\
Consejo Nacional de Investigaciones Cient\'ificas y T\'ecnicas
(CONICET), Argentina.}

\begin{abstract}
We extend some previous attempts to explain the origin and
evolution of primordial magnetic fields during inflation induced
from a 5D vacuum. We show that the usual quantum fluctuations of a
generalized 5D electromagnetic field cannot provide us with the
desired magnetic seeds. We show that special fields without
propagation on the extra non-compact dimension are needed to
arrive to appreciable magnetic strengths. We also identify a new
magnetic tensor field $B_{ij}$ in this kind of extra dimensional
theories. Our results are in very good agreement with
observational requirements, in particular from TeV Blazars and CMB
radiation limits we obtain that primordial cosmological magnetic
fields should be close to scale invariance.
\end{abstract}

 \maketitle

\section{Introduction}

Magnetic fields seem to be ubiquitous in the universe.
Observations have well established the widespread presence of
magnetic fields in the universe \cite{1988A&A...190...41K,
1989Natur.341..720K,1994RPPh...57..325K,2002ChJAA...2..293H,
2002ARA&A..40..319C,2004NewAR..48..763V,2005A&A...444..739B}.
Cosmic magnetism has been verified of $\mu$G-order strength in
galaxy clusters and also in high redshift protogalactic
structures. Recently, in particular, Kronberg~et~al and
Bernet~et~al reported organized, strong $B$-fields in galaxies
with redshifts close to 1.3~\cite{2008ApJ...676...70K,
2008Natur.454..302B}.  All these seem to suggest that magnetic
fields similar to that of the Milky Way are common in remote,
high-redshift galaxies. This could imply that the time needed by
the galactic dynamo to build up a coherent $B$-field is
considerably less than what is usually anticipated. On the other
hand, the widespread presence of magnetic fields at high redshifts
may simply mean that they are cosmological (pre-recombination) in
origin. Although it is still too early to reach a conclusion, the
idea of primordial magnetism gains ground, as more fields of
micro-Gauss strength are detected in remote proto-galaxies. Recent
reviews about primordial magnetogenesis can be found at
\cite{Kandus Report}.

Our work focuses in studying the production of primordial magnetic
seeds during an inflationary period, which in turn is induced by
the immersion of a particular 4D hypersurface, given by a de
Sitter spacetime, in a 5D vacuum space defined on a 5D
Riemann-flat extended de Sitter metric. The theory that motivates
such scenario is the Induced Matter Theory (IMT)\cite{IMT,imt1}.
In this theory inflation can be recovered from the
Campbell-Magaard
theorem\cite{campbell,c1,campbellb,campbellc,campbelld}, which
serves as a ladder to go between manifolds whose dimensionality
differs by one. This theorem, which is valid in any number of
dimensions, implies that every solution of the 4D Einstein
equations with arbitrary energy momentum tensor can be embedded,
at least locally, in a solution of the 5D Einstein field equations
in vacuum. Because of this, the stress-energy may be a 4D
manifestation of the embedding geometry. Physically, the
background metric there employed describes a 5D extension of an
usual de Sitter spacetime, which is the 4D spacetime that
describes an inflationary expansion.

We will study the simplest inflationary model, that gives us a de
Sitter epoch. Electromagnetic studies in this context have been
developed previously by \cite{2007Dahia, 2008Romero, 2005Liko}. To
introduce the electromagnetic effects we consider a massless
vector field $A_a$ of five components. The component normal to the
hypersurface has properties similar to a scalar field, so its
spectrum, with the Coulomb Gauge, we will see it can only be scale
invariant.

In the present paper we consider two types of fields. The first
with propagation outside the hypersurface, and the second with the
propagation confined to the hypersurface. This last fields need to
be generated by some brane source since they produce a
discontinuity in the stress tensor. Such distinction is very
important since the fields that can propagate outside the
hypersurface are strongly suppressed than the usual 4D photon,
giving rise to a bluer spectrum. The other fields, in contrast,
can give invariant spectrum for magnetic fields. Our results in
this way are very similar to \cite{2008JCAP...01..025M}.

Additionally, we analyze the electromagnetic fields measured by 4D
observers in the effective 4D hypersurface. When we extend our
definitions of the relativistic electric and magnetic fields to
more dimensions, we obtain new gravitoelectromagnetic fields of
scalar and tensorial nature. In our knowledge this is the first
time such components are identified in an electromagnetic theory
of Brane Worlds or Induced Matter Theory. Finally, it is well
known that the spatially flat FRW universe is conformal flat and
the Maxwell theory is conformal invariant, so that magnetic fields
generated during inflation would come vanishingly small at the end
of the inflationary epoch. However, in spatially open FRW models
magnetic fields are sufficiently strong to seed the galactic
dynamos. The reason is that conformal flatness of the 4D
background metric is a global property in spatially flat FRW
spacetimes, but a local one in spatially open or closed
geometries\cite{b1,b2,b3}.

This is a very important problem of inflation to explain seminal
magnetic fields. The possibility to solve this problem relies in
produce non-trivial magnetic fields in which conformal invariance
to be broken. The conformal invariance of Maxwell's equations in
four dimensions can also be broken if an embedding into a higher
dimensional space-time with time-varying extra spatial dimensions
is considered. In particular, in Brane Worlds or Induced Matter
theory conformal invariance is naturally broken, which make
possible the super adiabatic amplification of electric and
magnetic field modes during the early inflationary epoch of the
universe on cosmological scales.

\section{Electrodynamics a Riemann-flat extended de Sitter spacetime}

We begin considering the action of a free abelian gauge vector
field in five dimensions\footnote{In our conventions indices
"A,B,C,..,H" run from $0$ to $4$, Greek  indices run from $0$ to
$3$ and latin indices "i,j,k,..." run from $1$ to $3$.}

 \beg\label{accion5D}
    S=-\fr{1}{4}\int d^5x\sqrt{{}^{(5)}g}F_{CD}F^{CD};
 \en
where $F_{CD}=\na_C A_D-\na_D A_C=\pa_C A_D-\pa_D A_C$ is the
antisymmetric field tensor. The Maxwell equations of motion in
five dimensions are
\begin{equation}\label{Maxwell5D}
 \na^C F_{CD}=0.
\end{equation}

We introduce these fields in a background 5D spacetime in vacuum
state. Using some ideas of the Induced Matter Theories we want to
obtain a hypersurface that undergoes an effective inflationary
period. In this sense we do a the semiclassical expansion of the
vector fields
 \begin{equation}
A^C=\bar{A}^C+\de A^C,
 \end{equation}
where the overbar symbolizes the 3D spatially homogeneous
background field consistent with the fixed background homogeneous
metric and $\de A^C$ describes the fluctuations with respect to
$\bar{A}^C$. A more detailed explanation can be found in
\cite{mb2}. This space will be a 5D extension of a 4D de Sitter
spacetime, which is very relevant to inflationary cosmology, and
is given by a 5D (canonical Riemann-flat) metric with a line
element\cite{Ledesma2003}
 \beg\label{coord1}
    ds^2=\psi^2dN^2-\psi^2e^{2N}d\mbf{r}^2-d\psi^2,
 \en
where $N$ is a dimensionless time like coordinate related with the
number of e-folds during a de Sitter inflationary expansion. The
3D space like cartesian dimensionless coordinates
$d\mbf{r}^2=dx_1^2+dx_2^2+dx_3^2$ when taking a constant foliation
coincide with the 3D space coordinates of the effective 4D
hypersurface. The extra (non compact) space-like coordinate $\psi$
has spatial dimension. Instead of these coordinates, we shall use
new conformal ones [see appendix (\ref{conf coord})], such that
 \begin{equation}\label{coord3}
  ds^2=b(w)^2\left[a(\eta)^2\left(d\eta^2-d\mbf{R}^2\right)-dw^2\right],
 \end{equation}
where the scale factors $b(w)=e^{H_0w}$ y
$a(\eta)=-\fr{1}{H_0\eta}$ are dimensionless. The components of
the coordinates $(\eta,\bR,w)$ have spatial dimensions. The
foliation $\psi=\psi_0$ corresponds with $w=w_0=0$ in these
coordinates, such that $b(w_0)=1$.

The volume factor of the manifold is
$\sqrt{{}^{(5)}g}=b^5a^4=\fr{e^{5H_0w}}{H_0^4\eta^4}$, and the non
zero connections are
 \begin{equation}\label{conex3}
  \Ga_{\mu\nu}^0=-\eta^{-1}\de_{\mu\nu},\hspace{1cm}\Ga_{AB}^4={e^{-2H_0w}H_0}g_{AB},\hspace{1cm}
  \Ga_{\mu4}^\nu=H_0\de_\mu^\nu.\hspace{1cm}
 \end{equation}
The equations (\ref{Maxwell5D}) in these coordinates are
\footnote{The unique nonzero background field is the inflaton
which drives inflation and has a constant expectation value on the
effective 4D de Sitter hypersurface\cite{mb2}. Hence, in the
following we shall denote the fluctuations $ A^C\equiv\de A^C$}
\begin{equation}\label{Maxwell5D_coord3_0}
  -\left[\eta^2\pa^2+\fr{1}{H_0^2}\left(\fr{\pa^2}{\pa w^2}+H_0
  \fr{\pa}{\pa w}\right)\right]A_0 +
 \eta^2\fr{\pa}{\pa \eta}\pa_i A_i +\fr{1}{H_0^2}\fr{\pa}{\pa \eta}
 \left(\fr{\pa}{\pa w}+H_0\right)A_4 =0,
 \end{equation}
\begin{equation}
  \left[\eta^2\left(\fr{\pa^2}{\pa \eta^2}-\pa^2\right)-\fr{1}{H_0^2}\left(\fr{\pa^2}{\pa w^2}+
  H_0\fr{\pa}{\pa w}\right)\right]A_j +
 \eta^2\pa_j\pa_i A_i +\fr{1}{H_0^2}\pa_j\left(\fr{\pa}{\pa w}+H_0\right)A_4 =0,
 \end{equation}
\begin{equation}
   \left(\fr{\pa^2}{\pa \eta^2}-\fr{2}{\eta}\fr{\pa}{\pa\eta}-\pa^2\right) A_4 +
 \fr{\pa}{\pa w}\pa_i A_i +\fr{\pa}{\pa w}\left(\fr{2}{\eta}-\fr{\pa}{\pa \eta}\right)A_0
 =0,
 \end{equation}
where $\pa^2$ is the ordinary Laplacian operator.

We adopt the usual Coulomb Gauge for the study of the
electromagnetic effects on the hypersurface: $A_0=0$ y $\pa_i
A^i=0$. The equation for the temporal component $A_0$ remains as a
constraint equation for $A_4$,
\begin{equation}
 \fr{\pa}{\pa \eta}\left(\fr{\pa}{\pa w}+H_0\right)A_4=0.
 \end{equation}
Notice that in a 5D Minkowskian metric: $dS^2 = \eta_{ab} dx^a
\,dx^b$, the connections vanish and the field equations remain
decoupled after the gauge choice. In this case the universe does
not expands so that the inflaton field $A^4=0$.

To continue we propose the ansatz
 \beg\label{ansatz}
 {A^C}(\eta,\bR,w)=e^{mH_0w} {A}_m^C(\eta,\bR),
 \en
where $m$ can be any complex number. In particular, to make a
Fourier expansion base functions, we shall consider the allowed
values $m=-1/2+ip$, where $p=P/H_0$ are the wave numbers of the
field in the $w$ direction per unity of $H_0$. The reason to make
it relies in that $\sqrt{g}=e^{5H_0w}\eta^{-4}$ and
$F_{AB}F^{AB}\propto e^{-4H_0w}$, so that we obtain $\int d\eta\,
\int d^3x\, dw \,e^{H_0w}(\cdots)$ in the action. This exponent
should be suppressed in order to address a correct Fourier
expansion in terms of $e^{iPw+i\bk\cdot\bR}$.

It is also important to consider particular solutions of
perturbative order, without the oscillatory regime with respect to
the extra coordinate. These solutions are described by the real
values of $m$. In any case the equations are
 \bega
 (1+m)\fr{\pa }{\pa\eta}A_{4m}=0, \label{4m} \\
 \left[\eta^2\left(\fr{\pa^2}{\pa \eta^2}-\pa^2\right)-m(m+1)\right]{A_j}_m +
 \left(m+1\right)\pa_j{A_4}_m=0, \label{4m1}\\
  \left(\fr{\pa^2}{\pa \eta^2}-\fr{2}{\eta}\fr{\pa}{\pa\eta}-\pa^2\right) {A_4}_m
  =0. \label{4m2}
 \ena
We notice from (\ref{4m}) that the only possible mode for $A_{4m}$
that can exist is $m=-1$. The others yield $\pa^2 A_4=0$, such
that $A_{4m}=0$ when we suppose the absence of non trivial 3D
surface conditions. For $m=-1$ the equations of motion are
 \bega
 \left(\fr{\pa^2}{\pa \eta^2}-\pa^2\right){A_j}_{m=-1} =0,\\
  \left(\fr{\pa^2}{\pa \eta^2}-\fr{2}{\eta}\fr{\pa}{\pa\eta}-\pa^2\right) {A_4}_{m=-1} =0.
 \ena
Furthermore, the solutions with $m\neq-1$ remain decoupled since
${A_4}_{m\neq-1}=0$,
 \begin{equation}\label{ecmovAj}
  \left[\eta^2\left(\fr{\pa^2}{\pa
  \eta^2}-\pa^2\right)-m(m+1)\right]{A_j}_m=0.
 \end{equation}

\section{Electric and magnetic fields in 5D}

We need to define the electric and magnetic fields for the
observers that belong to the hypersurface.

When we extend the electromagnetic theory to five dimensions the
number of independent components of the antisymmetric
electromagnetic tensor $F_{CD}$ is $5(5-1)/2=10$. But the
interesting fact is that the dual tensor is of third rank.
 \beg
    \cf_{CDE}=\fr{1}{2}\eta_{CDEAB}F^{AB},
 \en
where the space volume tensor
$\eta_{CDEAB}=\sqrt{g}\,\epsilon_{CDEAB}$ is of 5th rank. If we
want to define the electric and magnetic components seen by
observers, we can apply the same definitions as in 4D. In this way
we obtain a 5D vector electric field
\begin{equation}
 E_C=u^D F_{CD},
\end{equation}
and a 2-tensor antisymmetric magnetic field \beg
    B_{CD}=\fr{1}{2}\eta_{CDEAB}u^B F^{EA}.
 \en
In particular, for a 5D comoving observer defined by
$u^A=a^{-1}b^{-1}(1,0,0,0,0)$ [in conformal coordinates
(\ref{coord3})], we obtain the following expression for the
components of this antisymmetric magnetic tensor
 \bega
    B_{(DD)}&=&0,\\
    B_{0D}&=&0,\\
    B_{4i}&=&\fr{1}{2}\sqrt{g}\,\eps_{4ijk0}\,u^0 \, F^{jk},\\
    B_{ij}&=&\sqrt{g}\,\eps_{ijk40}\,u^0 \,F^{k4},\label{B_ij}
 \ena
where we have distinguish the temporal, the 3-spatial and the
extra coordinate of the hypersurface. The number of independent
components is six: 3 belong to $B_{4i}$ playing the role of the
usual 3-magnetic vector field; while the other 3 components belong
to a new 3D antisymmetric magnetic tensor.

The electric field is also decomposed for this observer as
 \beg
  E_0=0,\hspace{1cm}
  E_i=\fr{1}{a\,b}\pa_0A_i,\hspace{1cm}
  E_4=\fr{1}{a\,b}\pa_0 A_4,
 \en
there are 3 components $E_i$ with the same properties as the usual
electric field, and an extra component $E_4$ that we call the
scalar electric field. Then the 10 components of the tensor
$F_{ab}$ decompose into: 3 of the usual electric vector field, 3
of the usual magnetic vector field, 3 of a new 3D magnetic tensor
field $B_{ij}$ and 1 of a new electric scalar field $E_4$.


\subsection{Quantization in the Coulomb Gauge}

Once we apply the Coulomb Gauge the dynamical fields are $A_j$ y
$A_4$. To perform a canonical quantization we should impose
commutation relations between the field and its conjugate
canonical momenta. From the 5D action (\ref{accion5D}) we define
the conjugate momenta $ \pi^B\equiv\fr{\de S}{\de
{A_B}_{,0}}=\sqrt{g}g^{00}g^{BC}{A_C}_{,0}$. In the coordinates
$(\eta,\mbf{R},w)$, they take the form
\begin{equation}
  \pi^j=b(w)^3a(\eta)^2g^{ji}{A_i}_{,0}, \qquad\pi^4=b(w)a(\eta)^2{A_4}_{,0}.
 \end{equation}
Therefore, the commuting rules at equal times are
  \begin{eqnarray}
\left[A^4(\eta,\bR,w),\pi_4(\eta,\bR',w')\right]&=&
b^3a^2\left[A_4(\eta,\bR,w),\fr{\pa
A_4}{\pa\eta}(\eta,\bR',w')\right]
            =i\de^{(3)}(\bR-\bR')\de(w-w'), \label{conmescalar1} \\
\left[{A}^j(\eta,\bR,w),\pi_i(\eta,\bR',w')\right]&=&
 b^3a^2\left[{A}^i(\eta,\bR,w),\fr{\pa{A}_j}{\pa\eta}(\eta,\bR',w')\right]
  = i\de(w-w'){\de^{(3)}_{\pe}}_j^i(\bR-\bR'), \label{conmvector1}
 \end{eqnarray}
where $${\de^{(3)}_{\pe}}_j^i(\bR-\bR) \equiv\int
\fr{d^3k}{(2\pi)^3}e^{i\bk\cdot(\bR-\bR')}\left(
\de_j^i-\de_{jl}\fr{k^ik^l}{k^2} \right),$$ is the 3D transversal
Dirac Delta distribution, which takes into account only
wavenumbers which are perpendicular to the direction of
propagation in order to obtain a correct quantization in the
Coulomb Gauge.

\subsection{5D Fourier decomposition}

The 5-vector $A^C$ has five polarization states. Considering
solutions from (\ref{Maxwell5D}) that propagate in all directions
 \begin{equation}\label{SoluA_i}
  {A_C}(s,\bk,p|\eta,\bR,w)=\eps_{Cs}(k,p)\ca_{s,k,p}(\eta)e^{i\bk\cdot\bR+ipH_0w},
 \end{equation}
where $s$ labels the polarization states, $k$ is the comoving
wavenumber in the 3D-spatial space of the hypersurface and $pH_0$
is the wavenumber in the extra direction $w$. To accomplish with
the Coulomb Gauge requirement we introduce the next orthonormal
base of polarization vectors:
  \begin{equation}
  \eps_0^A=b^{-1}a^{-1}( 1,\vec{0},0),\hspace{0.5cm} \eps_\la^A=b^{-1}a^{-1}(0,\vec{\eps}_\la,0),
\hspace{0.5cm}\eps_3^A=b^{-1}a^{-1}\left(0,\fr{k^i}{k},0
\right),\hspace{0.5cm}\eps_4^A=b^{-1}(0,\vec{0},1).
 \end{equation}
Here, $\la=1,2$ represents the transversal polarizations to the
3D-spatial propagation of the wave, they yield
$\vec{\eps}_\la\cdot\vec{\eps}_\la=1$ (no summing over $\la$).
With this choice the transversal vectors yield
$\eps^A_\la\eps_{A3}=\vec{\eps}_\la\cdot\bk=0$, that it is the
Coulomb Gauge in momentum $k$ space . The factor $b^{-1}a^{-1}$ is
important to define a vectorial base. The completeness relation
derived is
 \begin{equation}
  \sum_{s=0}^4 h^{ss'}\eps_s^A\eps_{s'}^B=g^{AB},
 \end{equation}
where the factor $h^{ss'}$ is matrix with the components of the 5D
Minkowski tensor metric. The previous expression projected to the
3D space is
 \begin{equation}
  \sum_{\la=1}^2\eps_\la^i(\bk)\eps_{j\la}(\bk) +\de_{jl}\fr{k^lk^i}{k^2}=\de_j^i.
 \end{equation}
This expression is very important to build the commutation rules.
Finally, the projection in the extra coordinate simply yields
 \begin{equation}
  h^{44} \,\eps_{s=4}^4\eps_{s=4}^4=g^{44}=-e^{-2H_0w}.
 \end{equation}

Using the last considerations, we expand the field
$A_i(\eta,\bR,w)$ in the solutions (\ref{SoluA_i})
 \bega \nonumber
  {A}_j(\eta,\bR,w)=e^{-\fr{H_0w}{2}}\int\fr{d^3k}{(2\pi)^{3/2}}\fr{dp}{(2\pi)^{1/2}}
  &&\sum_{\la=1}^2\eps_{j\la}(\bk,p)
  \left[ b_\la(\bk,p)e^{ipH_0w+i\bk\cdot\bR}{A}_{k,p}(\eta)\right.\\
  &&\left.+ b_\la^\dagger(\bk,p)e^{-ipH_0w-i\bk\cdot\bR}{A}^\ast_{k,p}(\eta)\right].
 \ena
The additional factor $e^{-H_0w/2}$ comes from the expansion in
functions $e^{ipH_0w+i\bk\cdot\bR}$ in such a way that the
integrand remains normalized.

Because of the factor $b^{-1}a^{-1}$ in the polarizations, the
function that yields the equation of motion (\ref{ecmovAj}) is:
$\ca_{k,p}(\eta)\equiv b(w)a(\eta){A}_{kp}(\eta)$. Then the mode
equation is
\begin{equation}\label{modosA1}
 \fr{d^2\ca_{kp}}{d\eta^2}+\left[k^2+\fr{\fr{1}{4}+p^2}{\eta^2}\right]\ca_{kp}=0.
\end{equation}
Notice that the value $m=-1/2+ip$ yields $-m(m+1)=|m|^2=1/4+p^2$
for the solutions that propagate on the extra coordinate. The
creation and annihilation operators satisfy the usual relations
(\ref{conmvector1})
 \begin{eqnarray}
\left[b_\la(\bk,p),b^\dagger_{\la'}(\bk',p')\right]&=&
 \de^{(3)}(\bk-\bk')\de(pH_0-p'H_0)\de_{\la\la'}, \\
 \left[b_\la(\bk,p),b_{\la'}(\bk',p')\right]&=&
 \left[b_\la^\dagger(\bk,p),b^\dagger_{\la'}(\bk',p')\right]=0.
 \end{eqnarray}
Finally, the normalization condition for the temporal modes
$\ca_{kp}(\eta)$ that comes from the commutation relations is
 \begin{equation}\label{normA1}
  a^2 b^2 W[A_{kp}(\eta)]=i,
 \end{equation}
where $W[f]=ff'^\ast-f^\ast f'$ is the Wronskian and $(')$ denotes
the partial derivative with respect to the conformal time $\eta$.

We have seen that the only scalar mode that propagates is
$A_{4m=-1}$, this means that an analog quantization in functions
$e^{ipH_0w+i\bk\cdot\bR}$ won't work in this gauge for $A_4$. In
its place we suppose small perturbative fields that only propagate
in the 4D hypersurface, decaying exponentially outside. The origin
of such fields is not addressed in the present paper.

\subsection{Vector modes}

In the following we will distinguish between modes that propagate
in all space directions, including the normal to the hypersurface
$w$, where $m=-1/2+ip$, and modes that only behave like plane
waves in the hypersurface, with $m\,\eps \mathbb{R}$.

\subsubsection{Extra dimensional propagation of the modes}

The solution for (\ref{modosA1}) of the modes that propagate in
all directions is
\begin{equation}
 \ca_{kp}(\eta)={d_1}_{(k,p)}(k\eta)^{1/2}\tilde{\ch}^{(1)}_{p}(k\eta)+
 {d_2}_{(k,p)}(k\eta)^{1/2}\tilde{\ch}^{(2)}_{p}(k\eta),
\end{equation}
where
$\tilde{\ch}^{(1)}_{p}(k\eta)\equiv\left(\tilde{J}_p+i\tilde{Y}_p\right)(k\eta)$
and
$\tilde{\ch}^{(2)}_{p}(k\eta)\equiv\left(\tilde{J}_p-i\tilde{Y}_p\right)(k\eta)$
are the Hankel functions with imaginary order $\nu=ip$,
$\tilde{J}_p(x)$ and $\tilde{Y}_p(x)$ are the first and second
kind Bessel functions with imaginary order defined real in the
following way \cite{NIST} [the reader can see some properties of
the Bessel functions with imaginary order in the appendix
(\ref{bes})]:
 \bega
  \tilde{J}_p(x)\equiv \fr{1}{\cosh\left(\fr{p\pi}{2}\right)}{\rm Re}\left[J_{ip}(x)\right], \label{bs1}\\
  \tilde{Y}_p(x)\equiv \fr{1}{\cosh\left(\fr{p\pi}{2}\right)}{\rm
  Re}\left[Y_{ip}(x)\right].\label{bs2}
 \ena
Here, we take into account the real part of the usual
decomposition of the Bessel functions. The quadratic amplitude of
these modes in the large scales is
 \begin{equation}\label{amplA_kp}
  \ca_{kp}\ca_{kp}^\ast|_{IR}\simeq \fr{\eta}{p}\left[\tanh\left(\fr{p\pi}{2}\right)
                    \cos^2\left[p\ln\left( \fr{k\eta}{2}\right)-\ga_p \right]+
                    \fr{\sin^2\left[p\ln\left( \fr{k\eta}{2}\right)-\ga_p \right]}
                    {\tanh\left(\fr{p\pi}{2}\right)}\right].
 \end{equation}
In the figure (\ref{Fig2}) we show the amplitude of the modes in
the effective 4D hypersurface versus $p$ for two different
wavenumbers $k$, during the inflationary epoch. The amplitude is
not divergent for $p\rightarrow 0$, since in this limit case we
have $(\tanh p)_{{p\rightarrow 0}}\rightarrow p$,
 $(\cosh p)_{p\rightarrow 0} \rightarrow 1$ and
$(\sinh p)_{p\rightarrow 0} \rightarrow p$, so that
\begin{equation}
  \ca_{k0}\ca_{k0}^\ast|_{IR}\simeq \fr{\pi\eta}{2}\left[1+\fr{4}{\pi^2}
  \left[ \ln\left(\fr{k\eta}{2}\right)-\ga\right]^2\right],
 \end{equation}
where $\ga\simeq 0.5772..$ is the Euler constant. On the other
hand, when $p\gg2/\pi$ one obtains $(\tanh p)_{p\gg2/\pi}
\rightarrow 1$, while the quadratic sines and cosines have the
same argument reducing to unity. Finally, we obtain
 \begin{equation}
  \ca_{kp\gg2/\pi}\ca_{kp\gg2/\pi}^\ast|_{IR}\simeq \fr{\eta}{p}=\fr{a^{-1}}{pH_0}.
 \end{equation}
Other important quantity is the amplitude of temporal derivatives
of the modes. In the long wavelength limit it takes the form
\begin{equation}\label{amplA'kp}
  \ca'_{kp}\ca_{kp}^{'\ast}|_{IR}\simeq \fr{4p}{\eta}\left[\tanh\left(\fr{p\pi}{2}\right)
                    \sin^2\left[p\ln\left( \fr{k\eta}{2}\right)-\ga_p \right]+
                    \fr{\cos^2\left[p\ln\left( \fr{k\eta}{2}\right)-\ga_p \right]}
                    {\tanh\left(\fr{p\pi}{2}\right)}\right].
\end{equation}
In the figure (\ref{Fig3}) we show the amplitude of the
derivatives of the modes for two different wavenumbers $k$. When
$p=0$, we obtain
 \beg
    \ca'_{k0}\ca_{k0}^{'\ast}|_{IR}\simeq
    \fr{8}{\pi\eta}=\fr{8H_0}{\pi} a(\eta),
 \en
while for $p\gg2/\pi$, we get
 \beg
    \ca'_{kp}\ca_{kp}^{'\ast}|_{IR}\simeq
    \fr{4p}{\eta}=4pH_0\,a(\eta),
 \en
so that in both limits grows as $H_0\, a(\eta)$.

\section{Electrodynamics on a 4D hypersurface}

With the aim to study the effective 4D dynamics of the fields
without propagation in a de Sitter spacetime, we shall consider a
static foliation $w=w_0$ on the metric (\ref{coord3}). This
hypersurface is relevant to describe a de Sitter expansion of the
universe in the early inflationary epoch.

\subsection{Effective 4D quantization}

Once we make the foliation on the fifth coordinate $w=w_0$ and the
modes are quantized [see appendix (\ref{quant})],  we obtain the
modes $\ca_{km}(\eta)\equiv b(w)\,a(\eta){A}_{km}(\eta)$ that
yield the equations of motion (\ref{ecmovAj})
 \begin{equation}\label{modosA2}
 \fr{d^2\ca_{km}}{d\eta^2}+\left[k^2-\fr{m(m+1)}{\eta^2}\right]\ca_{km}=0,
\end{equation}
which comply with the normalization condition
\begin{equation}\label{normA2}
 b^{2}\,a^2\,W[A_{km}(\eta)]=i,
\end{equation}
that is compatible with the commutation relationships
(\ref{conmvector2}).

\subsection{Inflaton modes on 4D}

Once fixed the Coulomb Gauge, and with our ansatz (\ref{ansatz}),
there is only one possible solution for the extra component $A_4$:
the one with the value $m=-1$. We call this solution as $\phi_k$,
that behaves as a massless scalar field in 4D. Using the equation
(\ref{modosA2}) we obtain the general solution
\begin{equation}
  \phi_k(\eta,w)=b^{-1}(k\eta)^{3/2}\left[c_1\ch_{\fr{3}{2}}^{(1)}(k\eta)+
  c_2\ch_{\fr{3}{2}}^{(2)}(k\eta)\right].
\end{equation}
From the condition (\ref{normA2}) we can set one constant to zero
$c_2=0$, so that $c_1(k)=\sqrt{\fr{H_0^2}{\pi k^3}}$. In the
infrared limit we obtain
 \begin{equation}
 \left. \phi_k\right|_{IR} =\fr{H_0}{\pi b}\left[\fr{4}{9}k^{\fr{3}{2}}\eta^3+i\fr{3}{2}k^{-\fr{3}{2}} \right].
 \end{equation}
This spectrum is exactly scale invariant, something we would
expect because we are dealing with 4D hypersurfaces $w=0$ that
suffer a de Sitter expansion. In this case the amplitude of the
modes, $|\phi_k|\equiv \sqrt{\phi_k \phi^*_k}$, is constant
 \begin{equation}\label{A_4(IR)}
  |\phi_k|\simeq\fr{3H_0}{2\pi b}k^{-\fr{3}{2}},
 \end{equation}
 in agreement with what one expects in a de
 Sitter expansion for a
 massless scalar field. In this sense this special solution has
 the appropriate spectrum of the inflaton fluctuations.

\subsection{Vector modes confined to the 4D hypersurface}\label{Propagacion 4D}

\begin{center}
\begin{figure}\begin{center}
  \includegraphics[width=9.5cm]{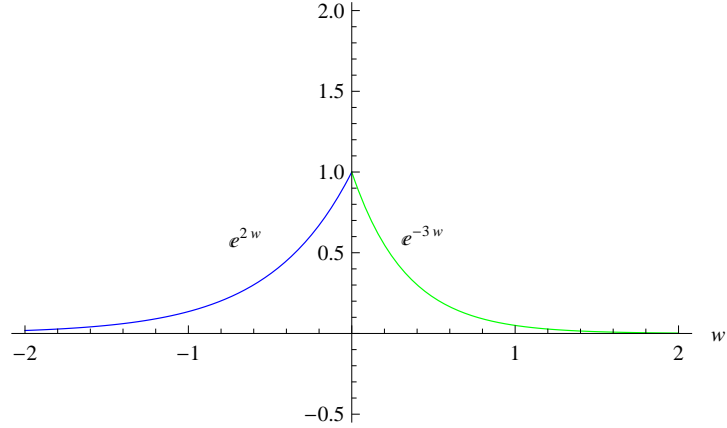}\\
\caption{Dependence of the modes outside the effective 4D
hypersurface using the shift symmetry
$m\leftrightarrow-(m+1)$}\label{Fig5}\end{center}
\end{figure}
\end{center}

Lets turn to study fields where their modes are confined to
propagate in the effective 4D hypersurface, which describes a
de
Sitter expansion for the universe. We proposed an ansatz for
such fields as ${A^C}(\eta,\bR,w)=e^{mH_0w} {A}_m^C(\eta,\bR)$,
depending exponentially in a parameter $m\in\mathbb{R}$. This
choice, however, can be problematic when the parameter $m<0$. In
this case the fluctuations blow when $w\rightarrow-\infty$. In the
same way, if $m>0$ the fluctuations blow as $w\rightarrow\infty$.
To avoid this problem we can suppose that these fluctuations reach
a maximum in the 4D hypersurface, and therefore decay
exponentially at both sides. A possible symmetry to choose is
$\mathbb{Z}_2$, common in brane theories. In this sense, for
example, if $m<0$, we have the solution $e^{-mH_0w}$ for $w<0$.
However, the solution will be $e^{mH_0w}$ for $w>0$. In this way
the function is continuous in $w=0$, but not its derivative that
suffers a discontinuity $2mH_0$. Apart from this, it appears a
complication since the equations of motion (\ref{modosA2}) change
when we go from one side to the other of the hypersurface. This is
due to the fact they are not invariant under the transformation
$m\leftrightarrow-m$. A closer inspection shows us that they are
invariant under the change $m\leftrightarrow-(m+1)$. This means
that the $m$-mode is equivalent to the $-(m+1)$-mode. Then, we use
the solution $e^{-(m+1)H_0w}$ when $m<-1$ for the interval $w<0$
and $e^{mH_0w}$ for the interval $w>0$. On the other hand, when
$m>0$ the solutions should be $e^{mH_0w}$ for $w<0$ and
$e^{-(m+1)H_0w}$ for $w>0$. This idea is sketched in the figure
(\ref{Fig5}). In the interval $-1<m<0$ we don't have this symmetry
and the solutions would irremediable blow. For this reason they
remain excluded.
The solution written in terms of Bessel functions, reads
 \begin{equation}
  \ca_{km}(\eta)=(k\eta)^{\fr{1}{2}}\left[{c_1}_{(k,m)}
   J_{\fr{1}{2}+m}(k\eta)+{c_2}_{(k,m)}J_{-\fr{1}{2}-m}(k\eta)  \right].
 \end{equation}
The quantum limit of this modes is
 \begin{equation}
\ca_{km}|_{UV}\longrightarrow\fr{1}{\sqrt{2k}}e^{-ik\eta},
\hspace{1cm} k\eta\longrightarrow -\infty.
 \end{equation}
 Using the normalization condition and the asymptotic form of the
Bessel functions we find
 \bega
 {c_1}_{(k,m)}&=&-{c_2}_{(k,m)}e^{-i\pi\left(\fr{1}{2}+m\right)},\\
 {c_2}_{(k,m)}&=& \sqrt{\fr{\pi}{4k}}\fr{e^{-i\pi\fr{(1-m)}{2}}}{\cos(m\pi)}.
 \ena

Once defined the modes for all scales we can find their large
scale limit for $(k\eta\rightarrow0)$
 \bega\label{A_km(IR)}
  \ca_{km}(\eta)|_{IR}&=&\fr{\sqrt{\pi}}{2^{\fr{1}{2}-m}}
                \fr{e^{-i\pi\fr{m}{2}}}{\Ga\left(\fr{1}{2}-m\right)\cos(m\pi)}
                       k^{-\fr{1}{2}}(k\eta)^{-m}+\\
              &+&\fr{\sqrt{\pi}}{2^{\fr{3}{2}+m}}
                 \fr{e^{i\pi\fr{(m+1)}{2}}}{\Ga\left(\fr{3}{2}+m\right)\cos((m+1)\pi)}
                       k^{-\fr{1}{2}}(k\eta)^{m+1}.
 \ena
This is the long wavelength solution for the modes $m<-1$ when
$w>0$. It is also the solution for $m>0$ when $w<0$. It is
interesting that the solutions do not change when we transform
$-m\leftrightarrow(m+1)$, in this case the first term shifts to
the second and the second to the first. It is useful to introduce
the following dimensionless function \cite{2008JCAP...01..025M} to
calculate the amplitude of the modes.
 \begin{equation}
  \cf(\mu)\equiv\fr{\pi}{2^{2\mu+1}\Ga^2{(\fr{1}{2}+\mu)}\cos^2(\mu\pi)},
 \end{equation}
where
 \begin{equation}
  \mu=-m,\ \ m>-\fr{1}{2}\hspace{1cm} ó    \hspace{1cm}\mu=m+1,\ \ m<-\fr{1}{2}.
 \end{equation}
The function only diverges in $\mu=-1/2$, dividing the function in
two well behaved branches. Previously we noted that only the modes
with $m\leq-1$ and $m\geq0$ have the symmetry to avoid divergences
and infinities. This means that for this modes the function is
well behaved.
\begin{figure}\begin{center}
\includegraphics[width=10cm]{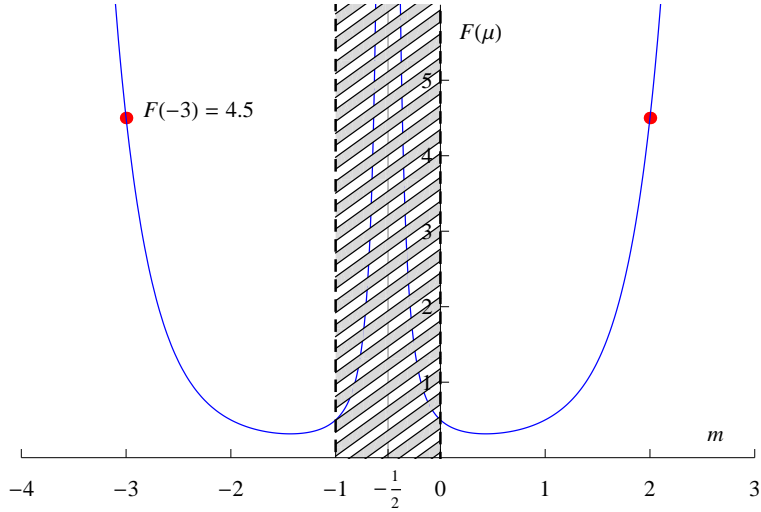}\\
\caption{The amplitude $\cf(m)$ for modes without propagation
outside the hypersurface. The invariant spectrum for magnetic
fields will be achieved for $m=-3$. The dashed sector is forbidden
so the fields are fluctuations that do not diverge in the extra
coordinates infinities.}\label{Fig6}
\end{center}\end{figure}

The amplitude of the modes in the hypersurface $w=0$ is
 \begin{equation}\label{amplA_km}
  \wh{\ca}_{k\mu}(\eta)\wh{\ca}_{k\mu}^\ast(\eta)|_{IR}\simeq k^{-1}\cf(\mu)(k\eta)^{2\mu}.
 \end{equation}
Another useful quantity is the temporal derivative of the modes
$\ca_{km}$. We find from (\ref{A_km(IR)})
 \begin{equation}
  \ca_{km}'(\eta)\simeq\sqrt{\pi}k^{\fr{1}{2}}ç
  \left[\fr{e^{-i\fr{m\pi}{2}}(-m)}{2^{\fr{1}{2}-m}
  \Ga\left(\fr{1}{2}-m\right)\cos(\pi m)}(k\eta)^{-(m+1)}
  +\fr{e^{i\fr{\pi (m+1)}{2}}(m+1)}{2^{\fr{3}{2}+m}
  \Ga\left(\fr{3}{2}+m\right)\cos(\pi (m+1))}(k\eta)^{m}\right].
 \end{equation}
Again we define a dimensionless function to describe the amplitude
of the modes
 \begin{equation}
  \cg(\theta)\equiv\fr{\pi(\theta+1)^2}{2^{3+2\theta}\,\Ga^2
  \left(\fr{3}{2}+\theta\right)\,\cos^2(\pi(\theta+1))},
 \end{equation}
where $\theta$ depends of $m$ in the following way
\begin{equation}
  \theta=-(m+1),\ \ m>-\fr{1}{2}\hspace{1cm} ó    \hspace{1cm}\theta=m,\ \ m<-\fr{1}{2}.
 \end{equation}
The amplitude reads
 \begin{equation}\label{amplA'_km}
  \ca_{k\theta}'\ca_{k\theta}'^\ast\simeq  k \cg(\theta) (k\eta)^{2\theta}.
 \end{equation}

\section{Electromagnetic energy densities}

The stress tensor we find from the action is
 \begin{equation}
  T_{AB}=-g^{CD}F_{AC}F_{DB}-\fr{1}{4}g_{AB}g^{CD}g^{EF}F_{DF}F_{CE}.
 \end{equation}
In the same way as in the usual 4D Maxwell theory, where
$T_{00}=-(E^2+B^2)$, we can introduce an expression for the
temporal-temporal component of the stress tensor as the sum of the
individual energies of each physical field
 \begin{equation}
  T_{00}=T_{00}^{E_i}+T_{00}^{B_i}+T_{00}^{E_4}+T_{00}^{B_{ij}}.
 \end{equation}
Lets write explicitly each of the previous contributions to the
energy density as seen by a 4D comoving observer. The vector
magnetic part of it, is
\begin{equation}
  {T^{(B_i)}}_{00}=-\fr{g_{00}}{2}g^{ij}g^{kl}F_{jl}.
 \end{equation}
Since $B_{4i}B^{4i}=g^{44}B_iB^i$ and $B_i=\eps_{ijk}a^{-1}\pa_j A_k$, we get
\begin{equation}
 B_{4i}B^{4i}=b^{-4}a^{-4}\left[\pa_jA_k \pa_jA_k-\pa_jA_k\pa_kA_j\right],
\end{equation}
so that
\begin{equation}
   {T^{(B_i)}}_{00}=-\fr{b^2a^2}{2}B_{4i}B^{4i}=\fr{1}{2}a^2 B_iB^i.
\end{equation}
This is the same result that obtained with the usual 4D
electrodynamics. The contribution due to the electric field
(vector) part is
\begin{equation}
   {T^{(E_i)}}_{00}=\fr{1}{2}g^{ij}F_{0i}F_{0j}=\fr{b^2a^2}{2}E_i\,E^i,
\end{equation}
and the electric (scalar) part
\begin{equation}
    {T^{(E_4)}}_{00}=\fr{b^2\,a^2}{2}E_4E^4.
\end{equation}
Finally, the contribution due to the tensor magnetic part is
\begin{equation}
    {T^{(B_{ij})}}_{00}=-\fr{b^2a^2}{2}B_{ij}B^{ij}.
\end{equation}

The energy density is defined as  $\rho\equiv-\lan {T}_0^0\ran$.
In our coordinates it takes the form
\begin{equation}
 \rho=\fr{1}{2}\left\lan-E_iE^i-E_4E^4+B_{4i}B^{4i}+B_{ij}B^{ij}\right\ran.
\end{equation}
All the terms are positive, $E_iE^i=-b^{-2}a^{-2}E_i^2$ and
$E_4E^4=-b^{-2}E_4^2$ [in comoving coordinates (\ref{coord3})].
The previous result in any coordinate system reads
 \beg
  \rho=\fr{1}{2}\left\lan-E_AE^A+B_{CD}B^{CD}\right\ran.
 \en

\subsection{Energy density due to Magnetic vector $B_i$}

We can calculate the contribution of the magnetic vector to energy
density defined as $\rho_{(B_i)}\equiv-\lan {T^{(B_i)}}_0^0\ran$:
 \beg
    \lan{T^{(B_i)}}_0^0\ran=-\fr{1}{2}\lan{B_{4i}B^{4i}}\ran.
 \en
Using the expansion in the modes $A_{kp}$, we obtain its
derivative in the direction $j_1$
 \beg
    \pa_{j_1}A_{j_2}=e^{-\fr{H_0w}{2}}\int\fr{d^3k}{(2\pi)^{\fr{3}{2}}}\fr{dp}{(2\pi)^{\fr{1}{2}}}
     \sum_{\la=1}^2 \eps_{\la j_2}   \left( (ik_{j_1})b_\la e^{i(\bk\cdot\bR+ipH_0w)}A_{kp}+
                  (-ik_{j_1})b_\la e^{-i(\bk\cdot\bR+ipH_0w)}A^\ast_{kp} \right).
 \en
Its vacuum expectation value is given by
 \begin{equation}
  \lan0|\pa_{j_1}A_{j_2}\pa_{j_3}A_{j_4}|0\ran=e^{-H_0 w}
  \int\fr{d^3k}{(2\pi)^{\fr{3}{2}}}\fr{dp}{(2\pi)^{\fr{1}{2}}} |A_{kp}|^2
   \sum_{\la=1}^2 k_{j_1}k_{j_3}\eps_{\la j_2}\eps_{\la j_4}.
 \end{equation}
So that $\lan {T^{(B_i)}}_0^0\ran$ can be written as
 \bega
 \nonumber \lan {T^{(B_i)}}_0^0\ran&=&-\fr{1}{2 \,b^5a^4}\int\fr{d^3k}{(2\pi)^3}
 \fr{dp}{2\pi}A_{kp}A_{kp}^\ast\\
            &&\times\sum_{\la=1}^2\left(  k_jk_j\eps_{i\la}\eps_{i\la}-
            k_i\eps_{i\la}k_j\eps_{j\la} \right).
 \ena
Using the completeness relation projected to the 3D space and the
transversal condition $k_i\vec{\eps_{i\la}}=0$, the second term in
the previous expression vanishes, while the first one holds
\begin{equation}
  \sum_{\la=1}^2\left(  k_jk_j\eps_{i\la}\eps_{i\la}-k_i\eps_{i\la}k_j\eps_{j\la} \right)=
  2k^2b^{2}a^{2}.
\end{equation}
Since the solution depends on the absolute value of $p$, we can write the integral as
$\int_{-\infty}^\infty dp=2\int_0^\infty dp$, so that the magnetic contribution to the
energy density is
 \bega
  \nonumber\rho_{(B_i)}&=&\fr{1}{2\pi^2 b^3}\int_0^\infty \fr{dk}{k}\fr{dp}{\pi}k^5
  \left|\fr{A_{kp}}{a}\right|^2\\
    &=&\fr{1}{2\pi^2b^5}\int_0^\infty \fr{dk}{k}\fr{dp}{\pi}k^5\fr{|\ca_{kp}|^2}{a^4}.
 \ena
The energy density stored at a certain scale $k$ for unity of $p$ is
 \begin{equation}
  \fr{d}{dkdp}\rho_{(B_i)}(\eta,k.p)=\fr{1}{2\pi^3}\fr{k^4}{b^5 a^4}\left|\ca_{kp}\right|^2.
 \end{equation}
Recalling (\ref{amplA_kp}) we arrive to
  \begin{equation}
  \fr{d}{dkdp}\rho_{(B_i)}(\eta,k,p)=\fr{1}{2\pi^3}\fr{k^4}{pH_0 b^5 a^5}
  \left[\tanh\left(\fr{p\pi}{2}\right)
                    \cos^2\left[p\ln\left( \fr{k\eta}{2}\right)-\ga_p \right]+
                    \fr{\sin^2\left[p\ln\left( \fr{k\eta}{2}\right)-\ga_p \right]}
                    {\tanh\left(\fr{p\pi}{2}\right)}\right].
 \end{equation}
The $k$-dependence of the energy density results of the integration of
$\fr{d}{dkdp}\rho_{(B_i)}(\eta,k,p)$ with respect to $p$. It is not possible to extend
to infinite values of $p$ when we perform the integration, this is because there is a
certain scale $p_0$ for which the modes don't have a quantum limit. This can be seen
from the mode equation (\ref{modosA1}), where the frequency is time dependent
\beg\omega^2_{kp}(\eta)=\fr{\fr{1}{4}+p^2}{\eta^2}+k^2.\en

This frequency is always positive, so the solutions oscillate, but depending on the
relative magnitudes between $p$ and the physical wavenumber $k_{phys}=k\,e^{-H_0t}$,
with respect to which of them oscillates.

In the previous asymptotic limits we have supposed that for modes in the ultraviolet
sector, and at the beginning of inflation $t=t_i$, we have $k_{phys}(t_i)\gg
H_0\sqrt{\fr{1}{4}+p^2}$ so $\omega_{kp}(t_i)\simeq k_{phys}(t_i)$. But, when inflation
ends at $t=t_f$, an interval of these wavelengths can increase such that
$k_{phys}(t_f)\ll p$ and then $\omega_{kp}(t_f)\simeq\sqrt{1/4+p^2}$. This means that
the actual scales of the observed universe $\la_{phys}$ will imply a cut $p_0$ in the
integral. The rest of the modes with $p>p_0$ do not have a quantum origin at the start
of inflation.

The connection of inflationary scales to the actual large scales
is something uncertain. It is known, approximately, the minimum
number of e-folds to solve the problems of planarity of the
Big-Bang, but this value depends strongly on the theoretical
model. Furthermore, during reheating there exists an important
re-scaling of the cosmological lengths \cite{2006Martin}.

A physical scale today at $t=t_h$, $\la^h_{phys}$ corresponds to
one at the start of inflation by
$\la^h_{phys}=\la^i_{phys}\fr{a_h}{a_i}$. In our case we are
dealing with a inflationary de Sitter expansion, where
$\fr{a_f}{a_i}=e^N$, such that the scales at the end of inflation
relate to the actual ones as
 \beg
    \fr{a_h}{a_f}=\fr{\sqrt{8\pi}}{R}\left[\fr{{\rho}_{inf}^2}
    {\Omega_{rad}^h\rho^h_{cri}}
    \right]^{1/4},
 \en
the Hubble parameter today is $H_h\simeq2.1h\times10^{-42}GeV$, the cosmological
radiation density parameter today is $\Omega^h_{rad}\simeq h^{-2}4.3\times10^{-5}$
\cite{Turner}, and the parameter $R$ that describes a phase of reheating
\cite{2006Martin} is in units of $m_{pl}$. It doesn't exist in the literature a model
in the framework of the Induced Matter theories that studies a realistic transition
between an inflationary period induced from a 5D vacua and a localized brane dominated
by radiation. We then assume that this transition is instantaneous. In our model we
obtain a vacuum energy density induced in the hypersurface
$$\rho_{inf}=\fr{3}{\kappa_5\psi_0^2}=\fr{3H_0^2}{\kappa_5}.$$
The critical energy density today is $$\rho_{cri}=\fr{3H_h^2 m_{pl}^2}{8\pi}.$$ The
parameter $R$ can be expressed as (for instantaneous reheating) $$\ln R=\fr{1}{4}\ln
\rho_{inf},$$ obtaining the following relation
 \beg
\fr{a_h}{a_f}=\left(\fr{\rho_{inf}}{\Omega_{rad}^h\rho^h_{cri}}\right)^{1/4}.
 \en
If we use the condition $k_{phys}(t_i)\gg H_0\sqrt{\fr{1}{4}+p^2}$ we can estimate a
cut for the values of $p$, so we are keeping modes with quantum initial conditions
 \beg
    p_0\ll\sqrt{\left[\fr{e^{2N}}{(\la^h_{phys})^2H_0H_h\sqrt{\Omega^h_{rad}}}\right]-\fr{1}{4}}.
 \en
There are three uncertainties in the last relation, the scale of
inflation $H_0/m_{pl}$ , the amount of inflation $N$ and the
scales of coherence of the cosmological magnetic field. Slight
variations in this parameters affect significatively the cut
$p_0$.
 In table \ref{table1} we give some examples on how these parameters $N$, $H_0$ y $\la^h_{phys}$
restrict $p_0$.
\begin{table}
\begin{center}
\begin{tabular}{|c|c|c|c|}
  \hline
  $N$ & $H_0$($m_{pl}$) & $\la^h_{phys}$(Mpc) & $p_0\ll$ \\
  \hline
  50 & $10^{-9}$ & $10^{4}$ & $-$ \\
  60 & $10^{-9}$ & $10^{4}$ & $45$      \\
  65 & $10^{-9}$ & $10^{4}$ & $6.7\times10^3$      \\
  50 & $10^{-6}$ & $10^{4}$ & $-$  \\
  60 & $10^{-6}$ & $10^{4}$ & $1.34$  \\
  65 & $10^{-6}$ & $10^{4}$ & $2.1\times10^2$  \\
  50 & $10^{-9}$ & $10^{2}$ & $-$      \\
  60 & $10^{-9}$ & $10^{2}$ & $4.5\times10^{3}$  \\
  65 & $10^{-9}$ & $10^{2}$ & $6.7\times10^{5}$  \\
  50 & $10^{-6}$ & $10^{2}$ & $-$  \\
  60 & $10^{-6}$ & $10^{2}$ & $1.4\times10^{2}$  \\
  65 & $10^{-6}$ & $10^{2}$ & $2.1\times10^{5}$  \\
  \hline
 \end{tabular}\caption{we show some relevant examples on how the parameters $N$, $H_0$ y $\la^h_{phys}$
restrict $p_0$.}\label{table1}
\end{center}
\end{table}

\begin{figure}
  \includegraphics[width=8cm]{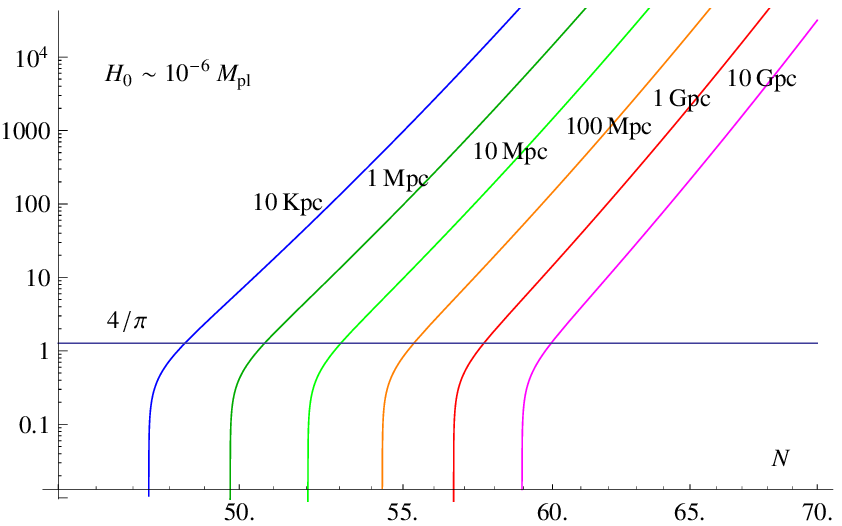}  \includegraphics[width=8cm]{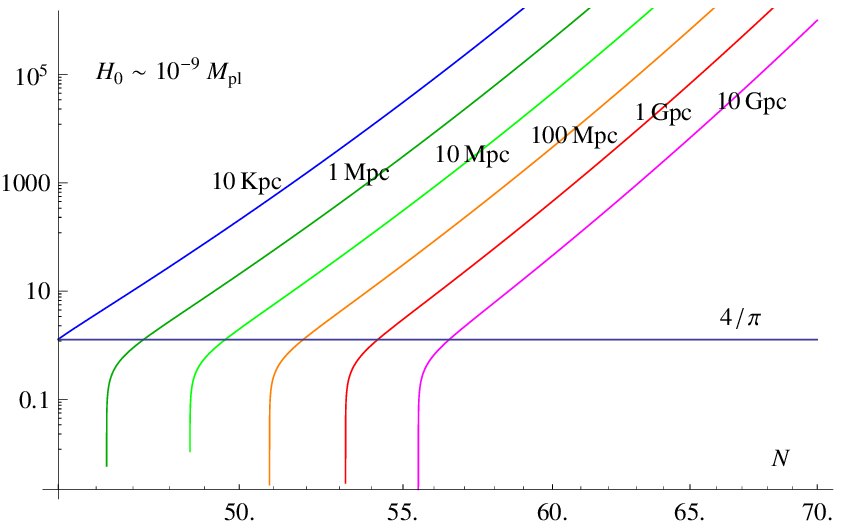}\\
\caption{The cut $p_0$ vs. the amount of inflation $N$ for small
proto-galactic scales of $10Kpc$ to very large scales of the
observable universe $10Gpc$. We have plotted two sub-Planckian
scales, in the left: $H_0\sim10^{-6}m_{planck}=10^{13}GeV$, and in
the right $H_0\sim10^{-9}m_{planck}$. It is important to notice
the change in the behavior of the cut at $4/\pi$} \label{p0}
\end{figure}

Using some numerical calculations [the reader can see the appendix
(\ref{apnum})], we can compute the magnetic energy density at a
certain scale $k$, for the modes that propagate on the extra
coordinate
$\fr{d}{dk}\rho_{(B_i)}(\eta,k,p_0)\equiv\int_0^{p_0}dp\fr{d}{dkdp}\rho_{(B_i)}(\eta,k,p)$
with
 \begin{equation}
  \fr{d}{dk}\rho_{(B_i)}(\eta,k,p_0)\simeq\fr{1}{2\pi^3}\fr{k^4}{H_0b^5 a^5}
  \ln\left(e^{1.303}\fr{p_0\pi}{8}\fr{k}{H_0\,a}\right),
 \end{equation}
where $p_0>4/\pi$. From the figure (\ref{p0}) we see that for
reasonable amounts of inflation, $N=60$, all of the relevant
scales are produced. However, for a lower value of e-folds,
$N=55$, and with a scale of inflation $H_0/m_{pl}\sim10^{-6}$
there are excluded the modes corresponding with scales larger than
$100\,Mpc$.

For the modes $\ca_{km}(\eta)$ we must compute
 \bega
  \nonumber\rho_{(B_i)}&=&\fr{1}{2\pi^2}\int_0^\infty \fr{dk}{k}\fr{k^5}{b^5} \left|\fr{A_{km}}{a}\right|^2\\
    &=&\fr{1}{2\pi^2}\int_0^\infty \fr{dk}{k}\frac{k^5}{b^5}\fr{|\ca_{km}|^2}{a^4}.
 \ena
The energy density at a certain scale $k$, is
 \begin{equation}
  \fr{d}{dln(k)}\rho_{(B_i)}(\eta,k)=\fr{1}{2\pi^3}\fr{k^5}{b^5\,a^4}\left|\ca_{km}\right|^2.
 \end{equation}
Recalling the amplitude (\ref{amplA_km}), we get
  \begin{equation}\label{densB}
  \fr{d}{dln(k)}\rho_{(B_i)}(\eta,k)=\fr{1}{2\pi^2}\fr{k^4}{b^5 a^4}\cf(\mu)(k\eta)^{2\mu}.
 \end{equation}
This also was obtained in \cite{2008JCAP...01..025M}, but in other
context.

\subsection{Energy density due to the magnetic tensor $B_{ij}$}\label{EnergiaBij}

The energy density that come from magnetic tensor is given by $\rho_{(B_{ij})}=-\lan
{T^{(B_{ij})}}_0^0\ran$
 \bega
  \rho_{(B_{ij})}=\fr{1}{2}\lan B_{ij}B^{ij}\ran= \fr{1}{2\pi^3}
  \int dkdp \fr{k^2}{b^5a^2}\left(\fr{1}{4}+p^2\right)H_0^2
  |\ca_{kp}|^2,
 \ena
from
$\fr{d}{dk}\rho_{(B_{ij})}(\eta,k,p_0)\equiv\int_0^{p_0}dp\fr{d}{dkdp}\rho_{(B_{ij})}(\eta,k,p)$
we obtain
 \begin{equation}
  \fr{d}{dk}\rho_{(B_{ij})}\left(\eta,k,p_0\right)\simeq\fr{1}{2\pi^3}\fr{k^2H_0}{b^5 a^3}\left[
  \fr{1}{4}\ln\left(e^{1.21}\fr{p_0\pi}{8}\fr{k}{H_0 a}\right)+\fr{p_0^2}{2}-\fr{8}{\pi^2}\right],
 \end{equation}
where $p_0>4/\pi$.

From the modes $\ca_{km}$ it is easy to see that
  \begin{equation}\label{bij}
  \fr{d}{dln(k)}\rho_{(B_{ij})}(\eta,k,\mu)=\fr{1}{2\pi^2}\mu^2H_0^2\cf(\mu)\fr{k^2}{b^5 a^2}(k\eta)^{2\mu}.
 \end{equation}
If $m=-1$ we also have a contribution from the scalar field $\phi_k$ in (\ref{A_4(IR)})
and after considering that $\cf(-1)=1/2$, we obtain
 \begin{equation}
  \fr{d}{dln(k)}\rho_{(B_{ij})}\left(\eta,k,m=-1\right)=\fr{13 H_0^2}{8\pi^2 b^5}\fr{k^2}{a^2}.
 \end{equation}

\subsection{Energy density due to the Electric vector $E_i$}

The energy density related to the electric field $E_i$ it is
 \bega
  \nonumber\rho_{(E_i)}=-\lan {T^{(E_i)}}_0^0\ran&=&\fr{1}{2a^4b^4}\lan\pa_0A_i\pa_0A_i\ran,\\
                     &=&\fr{1}{4\pi^2a^4b^5}\int dk\fr{dp}{\pi}k^2|\ca_{kp}'|^2.
 \ena
Calculating
$\fr{d}{dk}\rho_{(E_i)}(\eta,k,p_0)\equiv\int_0^{p_0}dp\fr{d}{dkdp}\rho_{(E_i)}(\eta,k,p)$
we get
 \begin{equation}
  \fr{d}{dln(k)}\rho_{(E_i)}(\eta,k,p_0)\simeq\fr{1}{\pi^3}
  \fr{k^3H_0}{b^5 a^3}\left[2.77+\fr{p_0^2}{2}\right],
 \end{equation}
with $p_0>4/\pi$.\\

The energy density associated to the modes $\ca_{km}$ is
 \begin{equation}
  \rho_{(E_i)}=\fr{1}{4\pi^2b^5 a^4}\int dk k^2|\ca_{km}'|^2
 \end{equation}
where we identify the energy density stored at a certain scale $k$
as
 \begin{equation}\label{ei}
  \fr{d\rho_{(E_i)}}{dln(k)}=\fr{1}{4\pi^2}\fr{k^4}{b^5a^4}\cg(\theta)(k\eta)^{2\theta}.
 \end{equation}
this is the same result derived in \cite{2008JCAP...01..025M}.

\subsection{Energy density due to the Electric scalar $E_4$}

The energy density corresponding to the scalar electric field is
$\rho_{(E_4)}=-\lan {T^{(E_4)}}^0_0\ran $, so that
 \bega
  \nonumber\rho_{(E_4)}&=&\fr{1}{2b^2a^4}\lan\pa_0 A_4\pa_0 A_4\ran,\\
  \nonumber      &=&\fr{1}{2b^3}\int \fr{dk}{2\pi^2}k^2\left|\fr{\phi_k'}{a}\right|^2,\\
                     &=&\int \frac{dk}{k} \fr{d\rho_{(E_4)}}{dln(k)},
 \ena
the energy density at a given scale $k$, is
 \begin{equation}
  \fr{d\rho_{(E_4)}}{dln(k)}=\fr{1}{(9\pi^2H_0)^2}\fr{k^6}{b^5a^6}.
 \end{equation}
this component decays strongly with the expansion. It is important
to notice that this contribution only appear when we are
considering a field that doesn't propagate in the extra dimension
and has a exponential decay in $w$ with $m=-1$.

\subsection{Spectrum of the fields}
  In table \ref{table2} we summarize the spectrums of the 5D electromagnetic fields for the
  two types of fields studied
 \begin{table}
\begin{center}
\begin{tabular}{|c|c|c|c|c|c|c|c|}
  \hline
  Power spectrum & 5D & $n$ & 4D & $m<-\fr{1}{2}$ & $m>-\fr{1}{2}$ &$\mu$&
   $n$ \\
  \hline &&&&\\
  $2\pi^2H_0^{-4}\fr{d}{\ln k}\rho_{(B)}$ & $x^5\ln\left(e^{1.303}\fr{p_0\pi}{8}x\right)$
  & $\sim5$ & & $\cf(m+1)x^{6+2m} $  & $\cf(-m)x^{4-2m}$   &$\cf(\mu)x^{4+2\mu}$ & $4+2\mu$ \\&&&&\\
  $2\pi^2H_0^{-4}\fr{d}{\ln k}\rho_{B_{ij}}$ & $x^3\ln\left(e^{1.21}\fr{p_0\pi}{8}x\right)$ &
  $\sim3$ & & $(m+1)^2\cf(m+1)x^{4+2m}$   & $m^2\cf(-m)x^{2-2m}$   &$\mu^2\cf(\mu)x^{2+2\mu}$ &$2+2\mu$ \\&&&&\\
  $2\pi^2H_0^{-4}\fr{d}{\ln k}\rho_{E_i}$ & $x^3\left(2.77+\fr{p_0^2}{2}\right)$ &
      $\sim3$ & &  $\cg(m)x^{4+2m}$   &   $\cg(-m-1)x^{2-2m}$   &$\cg(\theta)x^{4+2\theta}$ &$4+2\theta$ \\&&&&\\
  $9\pi^2H_0^{-4}\fr{d}{\ln k}\rho_{E_4}$ & --- & --- & $m=-1$ &$x^6$ & --- & --- &  $6$ \\&&&&
  \\
  \hline
\end{tabular}\caption{ The spectral indexes of the fields that can
propagate in the extra dimension are approximate, since there is a
logarithmic factor and so it doesn't strictly obey a power law.
This expressions for the energy density are very blue tilted, so
they decay to residual values at the end of inflation. The only
solutions that can give significant seeds for galactic magnetic
fields are does arising from real values of $m$. In particular,
when $m=-3$ we obtain an invariant spectrum for the vector
magnetic fields. For this value the vector electric and the tensor
magnetic fields are red tilted $n=-2$. For the fields with $m=-2$,
the electric and tensor magnetic fields are scale invariant, but
the vector magnetic fields is blue tilted with
$n_B=2$.}\label{table2}
\end{center}
\end{table}

\section{Back-reaction effects}

In order to make an analysis of the back-reaction effects we shall
estimate the energy density due to  $B_{ij}$ and the electric
(vector) field, which could be more problematic because, in both
cases their energy densities evolve as $(k\eta)^{-2}$. In order to
back-reaction effects to be negligible at the end of inflation, we
shall require that these energy densities to be smaller than the
critical background energy density: $\rho_{crit} =3 H_0^2
m_{pl}^2/(8\pi)$.

From the condition $\fr{d}{dln(k)}\rho_{(B_{ij})}<\rho_{crit}$ in
the eq. (\ref{bij}), we obtain that
\begin{equation}
\left(\frac{k}{a}\right)^{2(1+\mu)} < \frac{3 (H_0/m_{pl})^{2\mu}
}{4 {\cal F}(\mu) \mu^2} m_{pl}^{2(1+\mu)}.
\end{equation}
If we consider $\mu<-1$, modes for which back-reaction effects are
negligible has physical wavelengths of the order
\begin{equation}
\left.\lambda_{phys}\right|_{B_{ij}} < \left[\frac{4\mu^2 {\cal
F}(\mu)}{3}\right]^{\frac{1}{2(1+\mu)}}
\left(\frac{H_0}{m_{pl}}\right)^{\frac{1}{1+\mu}} \, \lambda_H.
\end{equation}
In order to make an estimation we can consider a scale invariant
magnetic spectrum with $\mu=-2$ and a Hubble parameter $H_0 =
1.0\times 10^{-5}\, m_{pl}$. In this case we obtain that the
\begin{equation}
\left.\lambda_{phys}\right|_{B_{ij}} < 0.204 \times
10^{5}\,\lambda_H,
\end{equation}
where $\lambda_H=1/H_0$ is the wavelength related to the Hubble
radius during inflation (we are considering $c=1$.

On the other hand, from the eq. (\ref{ei}), we obtain that the
condition to back reaction effects do not be relevant in the
background evolution of the universe, is (for cases with $\theta
<-2$)
\begin{equation}
\left.\lambda_{phys}\right|_{E_{i}} < \left[\frac{2 {\cal
G}(\theta)}{3 \pi}\right]^{\frac{1}{2(2+\theta)}}
\left(\frac{H_0}{m_{pl}}\right)^{\frac{1}{2+\theta}}\, \lambda_H,
\end{equation}
that for a scale invariant magnetic spectrum with $\mu=m=-2$, for
which $\theta=-3$, results to be
\begin{equation}
\left.\lambda_{phys} \right|_{E_{i}} < 0.512 \times 10^{5}\,
\lambda_H,
\end{equation}
which is bigger than the Hubble horizon during inflation. Hence,
in both cases, for $B_{ij}$ end $E_i$ the modes with wavelengths
which are on cosmological scales today (these wavelengths are
larger than the Hubble radius during inflation) are not
problematic in the back-reaction sense. On the other hand, it is
easy to see that $d\rho_{B_i}/[dln(k)]$ is or the order
$(H_0/m_{pl})^2 \sim 10^{-10}$ times smaller than the background
energy density, so that this contribution is negligible. Of
course, in our estimation we are considering a scale invariant
spectrum for the magnetic fields, but it can be seen from the
figure (\ref{FigB(varios)}) that (for $H_0 \simeq 10^{-5}\,
m_{pl}$), the admissible spectrums for the magnetic fields has a
spectral index $-0.12 < n_B < 0.1$ very close with an scale
invariant one. Our results in this sense are in agreement with
those obtained in \cite{2008JCAP...01..025M}. This topic has been
discussed in several articles\cite{a1,a2,a3}.

\section{Seed magnetic fields}

We are aimed to compare the present day strength of magnetic
fields with the predictions of our theory, we can estimate the
actual magnetic fields generated by this model, the magnetic
density parameter defined at a scale $L=2\pi/k$ is
 \beg
\Omega_B(k)\equiv\fr{\rho_{(B)}(t=t_h,k)}{\rho_{cri}}.
 \en
The energy density related to the magnetic fields after inflation
decay adiabatically with the expansion as $a^{-4}$
 \beg
    \rho_{(B)}(t=t_h,k)=\fr{d\rho_{(B)}}{d\ln k}(t=t_f,k)\left(\fr{a_f}{a_h}\right)^4.
 \en
Using the expression (\ref{densB}) for the magnetic density, we
obtain
 \beg
    \fr{d\Omega_{(B)}}{d\ln k}=\fr{32}{9}\cf(\mu)\left(\fr{\rho_{cri}}{m_{pl}}\right)
    \left(\fr{\rho_{cri}}{\rho_{inf}}\right)^\mu\left(\fr{a_h}{a_f}\right)^{2\mu}
    \left(\fr{k}{a_h\,H_h}\right)^{2\mu+4}.
 \en
Replacing the values for $\rho_{cri}$, $\rho_{inf}$ and
$(a_h/a_f)$ we may write \cite{2008JCAP...01..025M}
 \beg
 \fr{d\Omega_{(B)}}{d\ln k}\simeq 2.4\times10^{-7}(2.28\times10^{-58})^{\mu+2}\cf(\mu)R^{-2\mu}
 h^{2\mu+2}\left(\fr{k}{a_h\,H_h}\right)^{2\mu+4}.
 \en
Also, the magnetic density parameter is
 \beg
    h^2\fr{d\Omega_{(B)}}{d\ln k}\simeq 2.43\times \,10^{6}\left(\fr{B}{Gauss}\right)^2,
 \en
where we use that $1\,Tesla^2/2\simeq1.9\times10^{-32}\,GeV^4$ and
$1\,Tesla=10^4\,Gauss$. We then arrive to an expression for the seeds of the
cosmological magnetic field
 \beg\label{omb}
    \left(\fr{B}{Gauss}\right)\simeq\,3.16\times 10^{-7} h^{\mu+2}(2.28\times10^{-58})^{1+\fr{\mu}{2}}
    \cf(\mu)^{\fr{1}{2}}R^{-\mu}\left(\fr{k}{a_h\,H_h}\right)^{\mu+2}.
 \en
This is the final expression for magnetic fields generated from the fields $\ca_{km}$.
The values of $R$, without considering a particular model of reheating, are constrained
with a necessary cut for them to preserve nucleosynthesis:
$$\rho_{inf}>\rho_{nuc}\simeq 10^{-85}m_{pl}^4.$$ Furthermore, if inflation holds at
sub-Planckian scales we would need that
$H_{inf}/m_{pl}<1.3\times10^{-5}$, see \cite{2006Martin}. If
additionally we assume an instantaneous transition from reheating
to radiation [which is a good approximation on large
(cosmological) scales], and keeping in mind that in this case
$R^4\simeq (3/8\pi)(H_0^2\,m^2_{pl})$, we obtain that \beg
-24.7\lesssim\ln R\lesssim-5.71, \en which can be written in terms
of inflationary scales \beg
9\times10^{-43}<\fr{H_0}{m_{pl}}<1.3\times10^{-5}.\en The scale
invariant spectrum in (\ref{omb}) is achieved when
$m=-3$\footnote{For $m>-3$ the spectrum of magnetic fields is blue
tilted. In particular, for $m=-1$ is
\begin{displaymath}
\left(\fr{B}{Gauss}\right)_{m=-1}\simeq 5.09456\times
10^{-65}\,\left(\fr{k}{a_h\,H_h}\right)^{2}.
\end{displaymath}}
 \beg
    \left(\fr{B}{Gauss}\right)_{m=-3}\simeq8.00\times10^{-8}\left(\fr{H_0}{m_{pl}}\right).
 \en
The galactic dynamo scenario impose a lower limit for the seed
magnetic fields at the moment of the formation of seminal
galaxies, $B>10^{-22}Gauss$ (some models may relax to
$10^{-30}\,Gauss$). This then implies a constraining for the scale
of the inflation \beg\fr{H_0}{m_{pl}}>1.25\times10^{-15}.\en These
values are reasonable for sub-Planckian models. In particular, for
$H_0/m_{pl}\sim \,10^{-9}$, we obtain scale invariant cosmological
magnetic fields of amplitude
  \beg
    \left(\fr{B}{Gauss}\right)_{m=-3}\simeq \,8.00\times10^{-17},
 \en
which is compatible with dynamos mechanisms.

In the figure (\ref{FigB(m=-3)}) we plotted scale invariant
magnetic fields for different inflation scales $H_0/m_{pl}$ versus
wavelengths expressed in $Mpc$. The upper constraints come from
nucleosynthesis, Faraday Rotation measures and from the Cosmic
Microwave Background. Furthermore, we include the lower strengths
of magnetic fields needed to ignite the galactic dynamos and lower
magnetic fields from TeV Blazars observations. Notice that for
these spectrums the compatible scales are:
$4\,\times\,10^{-7}<H_0/m_{pl}<1.3\times10^{-5}$.

Very recently it was argued a lower bound $B\sim 3 \times
10^{-16}$ Gauss on the strength of intergalactic magnetic fields,
which stems from the un-observation of GeV gamma-ray emission from
electromagnetic cascade initiated by tera-electron volt gamma-ray
in an intergalactic medium from Fermi observations in TeV
Blazars\cite{nero}. More recently, it was argued that these
results are in tension with inflationary predictions
\cite{ponjas}. For our model this implies that the minimal scales
to produce fields of coherence bigger than $Mpc$ is $H_0\simeq
4\times 10^{-9}m_{pl}$. If the coherent fields are to be generated
for all scales, including smaller than $Mpc$ then the constraint
improves as $L^{-1/2}$ and the scale increases to
$4\times10^{-7}m_{pl}$.

\section{Conclusions}

Using some ideas of IMT, we have studied the magnetogenesis
produced by a 5D photon field in a hypersurface which undergoes a
period of de Sitter inflation. We deduce very blue tilted magnetic
fields for photons that can leak outside the 4D hypersurface, with
wavenumber $p H_0$. In order to produce significative seeds of the
magnetic fields to explain actual observations, we need other kind
of fields that cannot propagate outside the brane. These fields
produce a discontinuity in the stress tensor, so in turn their
origin should come from localized sources (i.e., static solutions)
on the extra space-like noncompact dimension. We do not study how
this fields came from, but once produced we analyze their
electromagnetic effects in the 4D effective de Sitter
(inflationary) universe. We have found that these fields are
candidates to produce relevant magnetogenesis. In particular, they
include the invariant scale spectrum for magnetic fields.

An important result here obtained is that we identify new physical
electromagnetic fields that arise from the extension of
Inflato-electromagnetic Inflation\cite{gi,mb1,mb2}, as an extended
Maxwell theory. Such fields are, for 4D comoving observes in the
hypersurface, a scalar electric field and an antisymmetric
magnetic tensor field. This formalism was proposed recently with
the aim to describe, in an unified manner, electromagnetic,
gravitational and the inflaton fields in the early inflationary
universe, from a 5D vacuum. Other conformal symmetry breaking
mechanisms have been proposed so far\cite{tur}. As in our case,
most of these are developed in the Coulomb gauge in order to
simplify the equations of motion for $A^{\nu}$. In our
calculations, the scalar electric field remains very constrained
by the actual model, since it is very blue tilted and exists for
one specific value of the parameter $m$. In contrast, the magnetic
tensor field, $B_{ij}$, and the vector electric components are red
tilted with respect to the vector magnetic field. Therefore, its
cosmological values at the end of inflation are very important.
The electric field is damped during reheating giving its energy to
charged particles. The tensor magnetic field should also be
extinguished, by transferring its energy to charged particles or
transforming to the vector magnetic field. The final fate of this
unobserved quantity depends on a model that should explain the
transition from this inflationary scenario to the actual universe.
However, this issue goes beyond the scope of this paper.

Finally, our results are in very good agreement with recent
observations\cite{nero}; if the coherent magnetic fields are to be
generated for all scales, including smaller than $Mpc$, magnetic
fields with strengths larger than $10^{-15}\, Gauss$ should be
generated during inflation on all scales for values of the Hubble
parameter larger than $H_0 > 4\times10^{-7}m_{pl}$, which is very
compatible with the accepted values during inflation. In the
figure (\ref{FigB(varios)}) we plotted different admissible
spectrums of magnetic fields. Notice that if we take into account
the more restrictive constrain that comes from TeV Blazars, we
obtain spectrums nearly scale invariants. The most blue tilted
compatible spectrum is for $n_B\simeq 0.1$ and a scale
$H_0=1.3\times 10^{-5}m_{pl}$. However, the most red tilted
magnetic fields take place for $n_B\simeq -0.62$ and a lower scale
$H_0\sim 10^{-15}m_{pl}$. Therefore, taking into account the
observational data the spectral indices for magnetic fields should
be in the range $-0.62 < n_B < 0.1$ for $ 10^{-15}\, m_{pl}< H_0 <
10^{-5}\, m_{pl}$.

\appendix

\section{Conformal coordinates}\label{conf coord}

 From (\ref{coord1}) with the transformation
 \beg
 t=\psi_0N,\hspace{1cm}\mbf{R}=\psi_0\mbf{r},\hspace{1cm}\psi=\psi,
 \en
we obtain the Ponce de Leon metric \cite{PdL}
 \beg\label{metricaPonce}
    ds^2=\left(\fr{\psi}{\psi_0}\right)^2\left[ dt^2-e^{2t/\psi_0}d\mbf{R}^2
    \right]-d\psi^2.
 \en
The foliations $\psi=\psi_0$ yield a 4D de Sitter hypersurface
with Hubble parameter $H=\psi_0^{-1}$.

The vacuum condition $R_{ab}=0$ is widely satisfied because
(\ref{metricaPonce}) is Riemann flat. On the other hand, the
transport of vectors in the space is not trivial due to the
existence of non zero connections. In coordinates (\ref{coord1})
the non null connections  are
 \begin{equation}
  \Ga_{\mu4}^\mu=\psi^{-1},\hspace{1cm}\Ga_{j0}^i=\de_j^i,\hspace{1cm}\Ga_{00}^4=\psi,
  \Ga_{ij}^0=e^{2N}\de_j^i,\hspace{1cm}\Ga_{ij}^4=-\psi e^{2N}\de_j^i.
 \end{equation}

Instead of working with the previous coordinates we will introduce new conformal like
coordinates. First we introduce a dimensionless coordinate
 \beg
    \cx=H_0\psi,
 \en
yielding
 \beg
    ds^2=\left(\fr{\cx}{\cx_0}\right)^2\left[ dt^2-e^{2H_0t/\cx_0}d\bR^2
    \right]-\fr{d\cx^2}{H_0^2}.
 \en
The foliation $\psi=\psi_0=H_0^{-1}$ corresponds in these
coordinates with $\cx=\cx_0=1$. Now we change to a conformal
coordinate
 \beg
  \mZ=\ln\cx, \hspace{1cm}d\mZ=\fr{d\cx}{\cx}=e^{-\mZ}d\cx,
 \en
such that the metric now reads
 \begin{equation}
  ds^2=e^{2\mZ}\left[e^{-\mZ_0}\left(dt^2-e^{2H_0e^{-\mZ_0}t}d\mbf{R}^2 \right)-
  \fr{d\mZ^2}{H_0^2}\right].
 \end{equation}
we can see that taking a constant foliation $\mZ=\mZ_1=cte$, we
obtain a 4D de Sitter space with Hubble parameter $H=H_0e^{\mZ_1}$
 \beg
  ds^2=dT^2-e^{2HT}d\mbf{r}^2,
 \en
where we have defined the cosmic time as $T=e^{\mZ_1-\mZ_0}t$. Here we can perceive
that the metric (\ref{coord1}) and their transformations generate $dS_4$-spaces for
constant foliations in the extra dimension.

To continue we define the usual conformal time $\eta=\int
a(t)^{-1}dt$ with $a(t)=e^{H_0e^{-\mZ_0}t}$. If we define the
coordinate $w=H_0^{-1}\mZ$, the metric finally reads
 \begin{equation}
  ds^2=b(w)^2\left[a(\eta)^2\left(d\eta^2-d\mbf{R}^2\right)-dw^2\right],
 \end{equation}
where the scale factors $b(w)=e^{H_0w}$ y
$a(\eta)=-\fr{1}{H_0\eta}$ are dimensionless. The components of
the coordinates $(\eta,\bR,w)$ have spatial dimensions. The
foliation $\psi=\psi_0$ corresponds with $w=w_0=0$ in these
coordinates, such that $b(w_0)=1$.

\section{Mode quantization}\label{quant}

In order to obtain the canonical structure of
${}^{(5)}A^4_{m}(\eta,\bR,w_0)$ and its canonical momentum, we
must calculate the effective 4D commutators
 \bega\label{conmescalar2}
\nonumber  \left[{}^{(5)}A^4_{m}(\eta,\bR,w_0),
\pi_{4m'}(\eta,\bR,w_0)\right]&=&
  b^3\,a^2\left[{}^{(5)}A^4_{m}(\eta,\bR,w_0),
  \fr{\pa}{\pa\eta}{}^{(5)}A^4_{m'}(\eta,\bR',w_0)\right]\\
  &=&i\de^{(3)}(\bR-\bR')\de_{mm'}b^{(1+m+m')}.
 \ena
The quantization follows in the usual way, by expanding in plane
waves $e^{i\bk\cdot\bR}$
\begin{equation}
  {}^{(5)}A_{4m}(\eta,\bR,w_0)=e^{mH_0w_0}\int\fr{d^3k}{(2\pi)^{3/2}}{\eps_4}_{s=4}\left[ a_{\bk m}
e^{i\bk\cdot\bR}\phi_{km}(\eta)+
              a^\dagger_{\bk m}e^{-i\bk\cdot\bR}\phi^\ast_{km}(\eta)\right].
 \end{equation}
The creation and annihilation operators yield the relations
\begin{equation}
  \left[ a_{\bk m},a_{\bk'm'}^\dagger \right]=\de^{(3)}(\bk-\bk')\de_{mm'},\hspace{1cm}
  \left[ a_{\bk m},a_{\bk'm'} \right]=\left[ a_{\bk m}^\dagger,a_{\bk'm'}^\dagger
  \right]=0.
 \end{equation}
The mode equation is
 \begin{equation}\label{modosphi}
  \fr{d^2{\phi}_{km}}{d\eta^2}-2\eta^{-1}\fr{d{\phi}_{km}}{d\eta}+k^2{\phi}_{km}=0,
 \end{equation}
and the normalization condition over them
\begin{equation}\label{normesca1}
 b^{2}a^2W[\phi_{km}(\eta)]=i,
 \end{equation}
such that the only accepted value is $m=-1$, for this model with the Coulomb Gauge
choice.

For the transversal vector field there is no restriction over the
values of $m$
 \bega\label{conmvector2}
\nonumber
\left[{}^{(5)}{A}^j_m(\eta,\bR,w_0),\left(\pi_{i}\right)_{m'}(\eta,\bR',w_0)\right]&=&
 b^3a^2\left[{}^{(5)}{A}^j_m(\eta,\bR,w_0),\fr{\pa}{\pa\eta}{}^{(5)}{A}_{im'}(\eta,\bR',w')\right]\\
&=&i\de_{mm'}e^{(1+m+m')H_0w_0}{\de^{(3)}_{\pe}}_j^i(\bR-\bR').
\ena

The mode expansion is given by
\begin{equation}
  {}^{(5)}A_{jm}(\eta,\bR,w_0)=e^{mH_0w_0}\int\fr{d^3k}{(2\pi)^{3/2}}
  \sum_{\la=1}^2{\eps_j}_{\la}(\bk m)
\left[ {b_\la}(\bk m)e^{i\bk\cdot\bR}
A_{km}(\eta)+{b^\dagger_\la}(\bk
m)e^{-i\bk\cdot\bR}A^\ast_{km}(\eta)\right],
 \end{equation}
where the creation and annihilation operators describe an algebra
\begin{equation}
 \left[b_\la(\bk,m),b^\dagger_{\la'}(\bk',m')\right]=
 \de^{(3)}(\bk-\bk')\de_{mm'}\de_{\la\la'},\hspace{1cm}
 \left[b_\la(\bk,m),b_{\la'}(\bk',m')\right]=
 \left[b_\la^\dagger(\bk,m),b^\dagger_{\la'}(\bk',m')\right]=0.
\end{equation}

\section{Some properties of the Bessel functions with imaginary
order}\label{bes}

\begin{figure}\begin{center}
  \includegraphics[width=9.5cm]{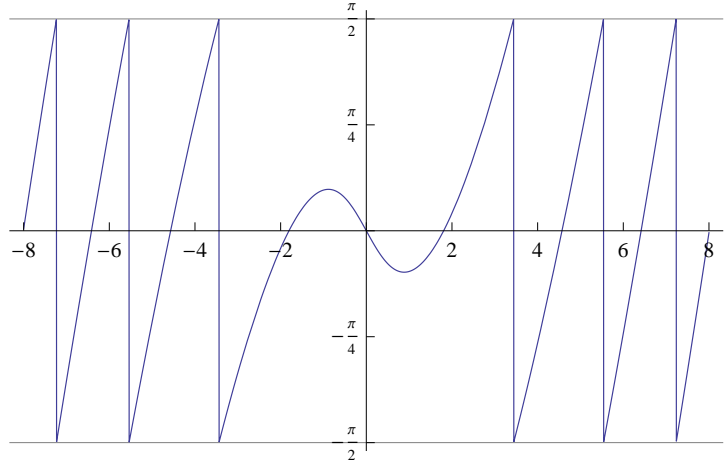}\\
  \caption{The phase of the complex function $\Ga(1+ip)$}\label{Fig1}
\end{center}\end{figure}

The decomposition of the functions (\ref{bs1}) and (\ref{bs2}) is
complex because of their imaginary order. To apply the
normalization condition we shall use the asymptotic behavior in
the short wavelength limit, which describes the modes in the UV
sector of the spectrum
 \bega
  \tilde{J}_p(x\rightarrow+\infty)\simeq \sqrt{\fr{2}{\pi x}}
  \cos\left(x-\fr{\pi}{4}\right)+\co(x^{-3/2}),\\
  \tilde{Y}_p(x\rightarrow+\infty)\simeq \sqrt{\fr{2}{\pi x}}
  \sin\left(x-\fr{\pi}{4}\right)+\co(x^{-3/2}).
 \ena
It is interesting to notice that in this limit the modes do not
depend in the order $p$, so in this limit these functions become
the usual Bessel functions with zero order
 \bega
 \tilde{\ch}_{ip}^{(1)}(x\rightarrow+\infty)\simeq\sqrt{\fr{\pi}{2x}}
 e^{-ix-i\fr{\pi}{4}}\simeq{\ch}_{0}^{(1)}(x\rightarrow+\infty),\\
 \tilde{\ch}_{ip}^{(2)}(x\rightarrow+\infty)\simeq\sqrt{\fr{\pi}{2x}}
 e^{+ix+i\fr{\pi}{4}}\simeq{\ch}_{0}^{(2)}(x\rightarrow+\infty).
 \label{bs}
 \ena
In the short wavelength limit, deep inside the horizon, the fluctuations travel like
plane waves in the Minkowski spacetime
\begin{equation}
  \ca_{kp}|_{UV}\longrightarrow\fr{1}{\sqrt{2k}}e^{-ik\eta}, \hspace{1cm} k\eta\longrightarrow -\infty.
\end{equation}
The Hankel function $\tilde{\ch}_{p}^{(1)}(x)$ has the desired asymptotic limit. This
means that we can set $d_2=0$ and using the normalization relationship (\ref{normA1}),
that in this limiting case is simply $W[\ca_{kp}(\eta)]=i$, we arrive to
$d_1=\fr{1}{\sqrt{k\pi}}$. The solution for all scales finally reads
\begin{equation}
 \ca_{kp}(\eta)=\sqrt{\fr{\pi\eta}{2}}\tilde{\ch}_{p}^{(1)}(k\eta).
\end{equation}
When the physical wavelengths are much bigger than the Hubble
horizon the Bessel functions of imaginary order has the following
asymptotic behavior
 \bega
  \tilde{J}_p(x\rightarrow0^+)&\simeq&\sqrt{\fr{2\tanh\left(\fr{p\pi}{2}\right)}{p\pi}}
  \cos\left[p\ln\left( \fr{x}{2}\right)-\ga_p \right],\\
  \tilde{Y}_\si(x\rightarrow0^+)&\simeq&\sqrt{\fr{2}{p\pi\tanh\left(\fr{p\pi}{2}\right)}}
  \sin\left[p\ln\left( \fr{x}{2}\right)-\ga_p \right],
 \ena
where $\ga_p$ is a phase \cite{NIST} defined through
 \begin{equation}
  \tan\ga_p=\fr{{\rm Im}[\Ga(1+ip)]}{{\rm Re}[\Ga(1+ip)]},
 \end{equation}
that complies with $\ga_0=0$. This phase was plotted in
(\ref{Fig1}).

\section{Numerical calculation of $\fr{d}{dk}\rho(\eta,k,p_0)$}\label{apnum}

Now we must resolve the integral
\begin{equation}
  \int_0^{p_0}\eta^{-1}|\ca_{kp}(\eta)|^2 dp=\int_0^{p_0}
  \fr{dp}{p}\left[\tanh\left(\fr{p\pi}{2}\right)
  \cos^2\left[p\ln\left( \fr{k\eta}{2}\right)-\ga_p \right]+
  \fr{\sin^2\left[p\ln\left( \fr{k\eta}{2}\right)-\ga_p \right]}
  {\tanh\left(\fr{p\pi}{2}\right)}\right].
\end{equation}
We shall resolve approximately this integral. To do it, we
separate it in two parts
$\int_0^{p_0}=\int_0^{4/\pi}+\int_{4\pi}^{p_0}$. From $p=4/\pi$
the modes are very close to their asymptotic form
 \beg
    \int_0^{p_0}\eta^{-1}|\ca_{kp}(\eta)|^2 dp=\int_0^{4/\pi}
  \fr{dp}{p}\left[\tanh\left(\fr{p\pi}{2}\right)
  \cos^2\left[p\ln\left( \fr{k\eta}{2}\right)-\ga_p \right]+
  \fr{\sin^2\left[p\ln\left( \fr{k\eta}{2}\right)-\ga_p \right]}
  {\tanh\left(\fr{p\pi}{2}\right)}\right]+\int_{4/\pi}^{p_0}\fr{dp}{p}.
 \en
The first integral can be numerically resolved
\begin{center}
 \includegraphics[width=10cm]{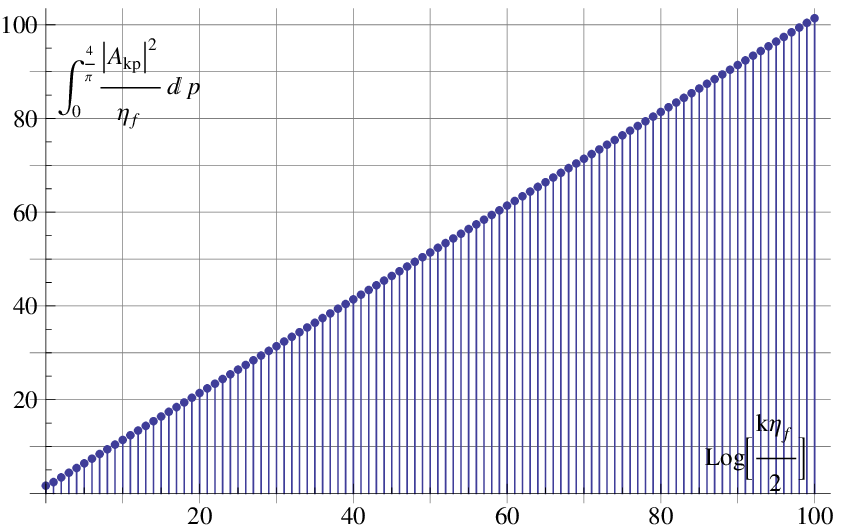}
\end{center}
We see that the function approximates linearly to
$\ln\left(\fr{k\eta}{2}\right)$. We get
\begin{equation}
  \int_0^{4/\pi}\eta^{-1}|\ca_{kp}(\eta)|^2 dp\simeq \ln\left(\fr{k\eta}{2}\right)+1.303.
\end{equation}
The complete integral is then
\begin{equation}
  \int_0^{p_0}\eta^{-1}|\ca_{kp}(\eta)|^2 dp\simeq \ln\left(e^{1.303}\fr{p_0\pi}{8}k\eta\right).
\end{equation}

The energy density that come from magnetic tensor is given by
$\rho_{(B_{ij})}=-\lan {T^{(B_{ij})}}_0^0\ran$
 \bega
  \rho_{(B_{ij})}=\fr{1}{2}\lan B_{ij}B^{ij}\ran= \fr{1}{2\pi^3}
  \int dkdp \fr{k^2}{b^4a^2}\left(\fr{1}{4}+p^2\right)H_0^2
  |\ca_{kp}|^2,
 \ena
so the energy at a certain scale $k$ with extra momentum $p$ is
 \begin{equation}
\fr{d\rho_{B_{ij}}}{dkdp}=\fr{1}{2\pi^3}\fr{k^2H_0^2
\left(\fr{1}{4}+p^2\right)}{b^5a^2}|\ca_{kp}|^2.
 \end{equation}
As before we split the integral in two parts and solve numerically
until the asymptotic regime is reached
\begin{center}
\includegraphics[width=10cm]{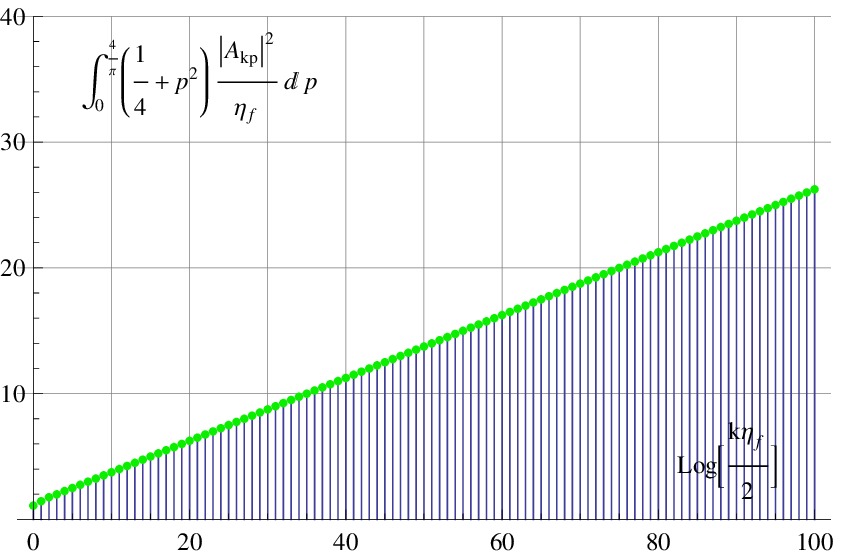}
\end{center}
the linear approximation gives
\begin{equation}
  \int_0^{4/\pi}\eta^{-1}\left(\fr{1}{4}+p^2\right)|\ca_{kp}(\eta)|^2 dp\simeq
  \fr{1}{4}\ln\left(\fr{k\eta}{2}\right)+1.21.
\end{equation}

The energy density related to the electric field $E_i$ it is
 \bega
  \nonumber\rho_{(E_i)}=-\lan {T^{(E_i)}}_0^0\ran&=&\fr{1}{2v^4u^4}\lan\pa_0A_i\pa_0A_i\ran,\\
                     &=&\fr{1}{4\pi^2a^4b^5}\int dk\fr{dp}{\pi}k^2|\ca_{kp}'|^2.
 \ena
This energy stored at a scale $k$ with extra-momentum $p$ is
 \begin{equation}
  \fr{d\rho_{(E_i)}}{dkdp}=\fr{1}{4\pi^3}\fr{k^2}{b^5a^4}|\ca_{kp}'|^2.
 \end{equation}
From the expression (\ref{amplA'kp}) for the amplitude of the
temporal derivatives of the modes we get
 \beg
\fr{d\rho_{(E_i})}{dkdp}=\fr{1}{\pi^3}\fr{pH_0k^2}{b^5a^3}\left[\tanh\left(\fr{p\pi}{2}\right)
                    \sin^2\left[p\ln\left( \fr{k\eta}{2}\right)-\ga_p \right]+
                    \fr{\cos^2\left[p\ln\left( \fr{k\eta}{2}\right)-\ga_p \right]}
                    {\tanh\left(\fr{p\pi}{2}\right)}\right].
 \en
The numeric integral to $p=4/\pi$ gives us
\begin{center}
  \includegraphics[width=10cm]{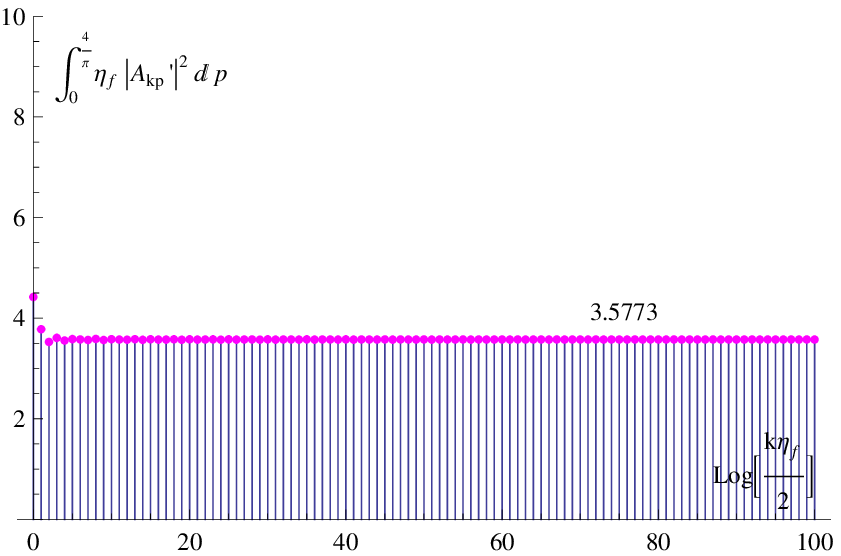}
\end{center}
that is a constant value.

\newpage

\begin{figure}
 \includegraphics[width=8cm]{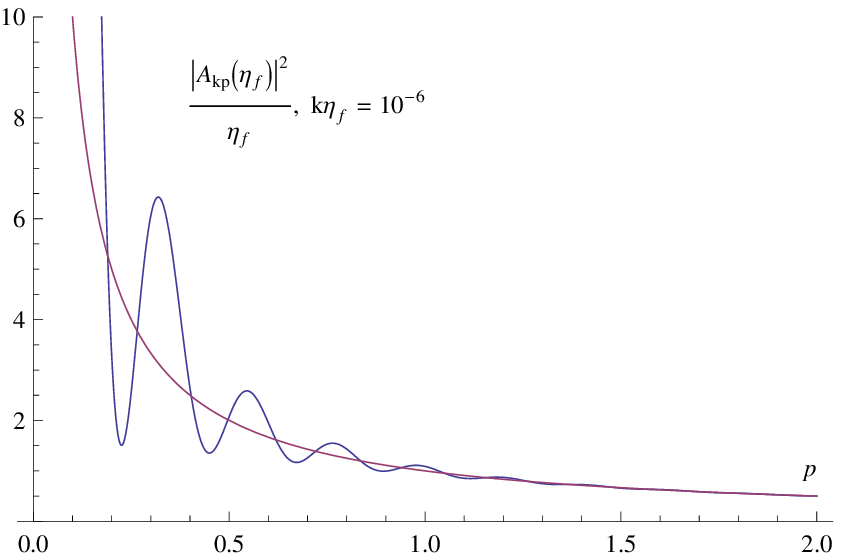}
 \includegraphics[width=8cm]{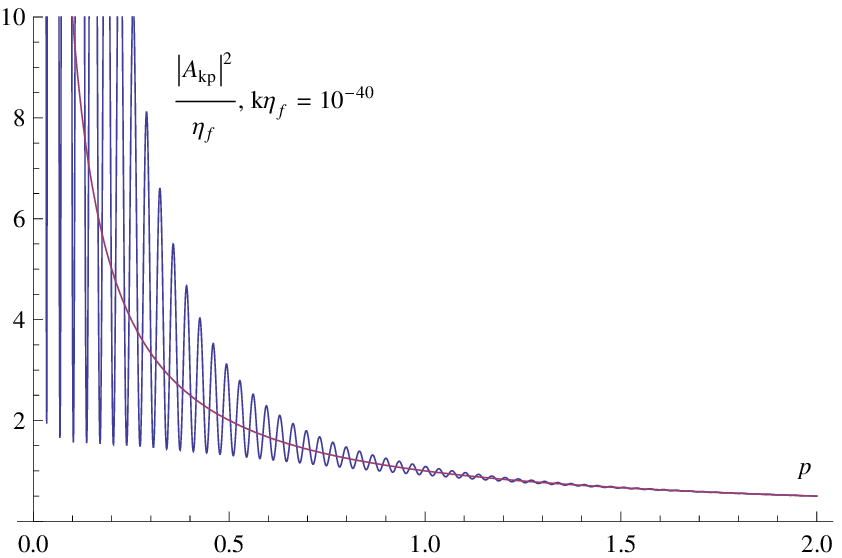}\\
\caption{The amplitude of the modes vs. $p$ for two different
scales during the inflationary period in the hypersurface. We can
see that the modes collapse to $p^{-1}$ for
$p>>2/\pi$.}\label{Fig2}
\end{figure}

\begin{figure}\begin{center}
\includegraphics[width=8cm]{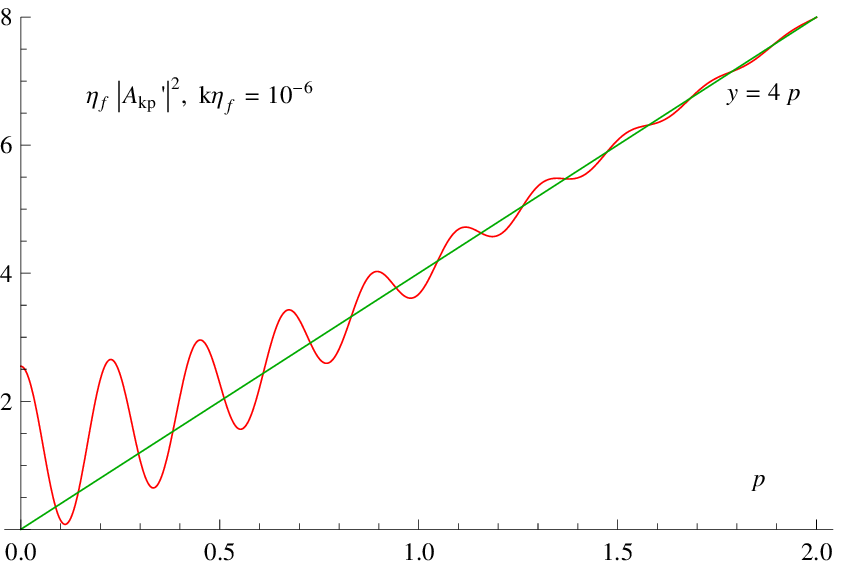}
\includegraphics[width=8cm]{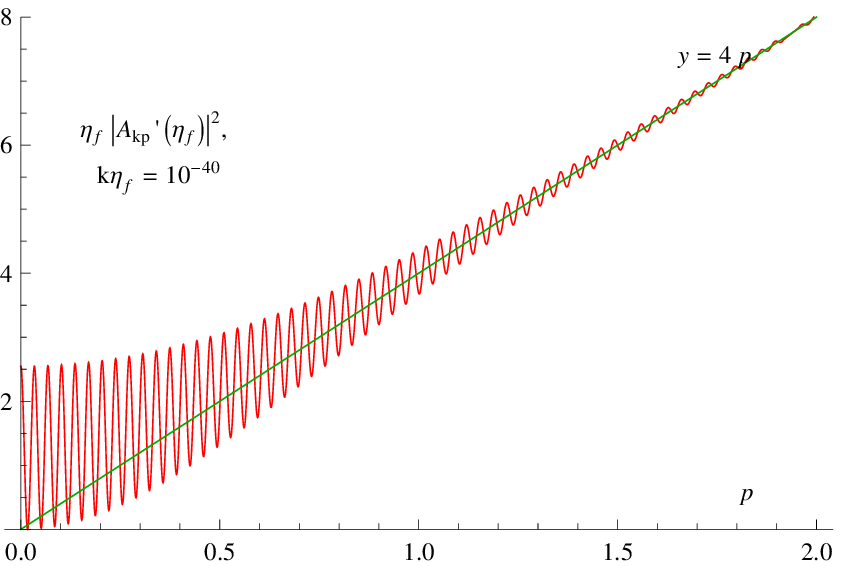}\\
\caption{Amplitude of the derivatives of the modes for two scales.
The asymptotic value is $4p$}\label{Fig3}\end{center}
\end{figure}
\begin{figure}
\includegraphics[width=12cm]{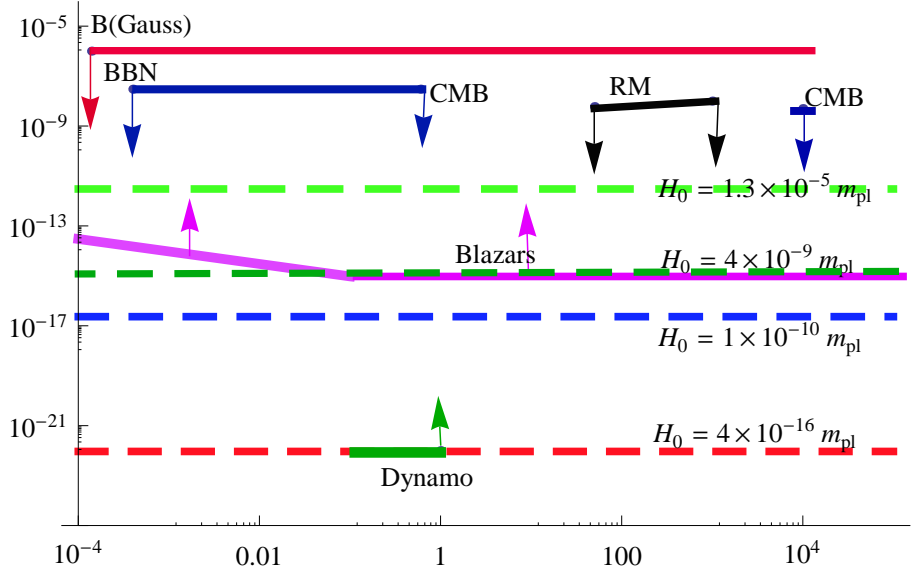}\\
\caption{Scale invariant magnetic fields for different inflation
scales $H_0/m_{pl}$ plotted against cosmic scales in $Mpc$. The
upper constraints come from nucleosynthesis, Faraday rotation
measures and from the Cosmic Microwave Background. We also include
the lower magnetic fields needed to ignite the galactic dynamos
and TeV Blazars. For these spectrums the compatible scales are: $4
\times 10^{-7}<H_0/m_{pl}<1.3\times10^{-5}$.}
  \label{FigB(m=-3)}
  \end{figure}
\begin{figure}
\includegraphics[width=12cm]{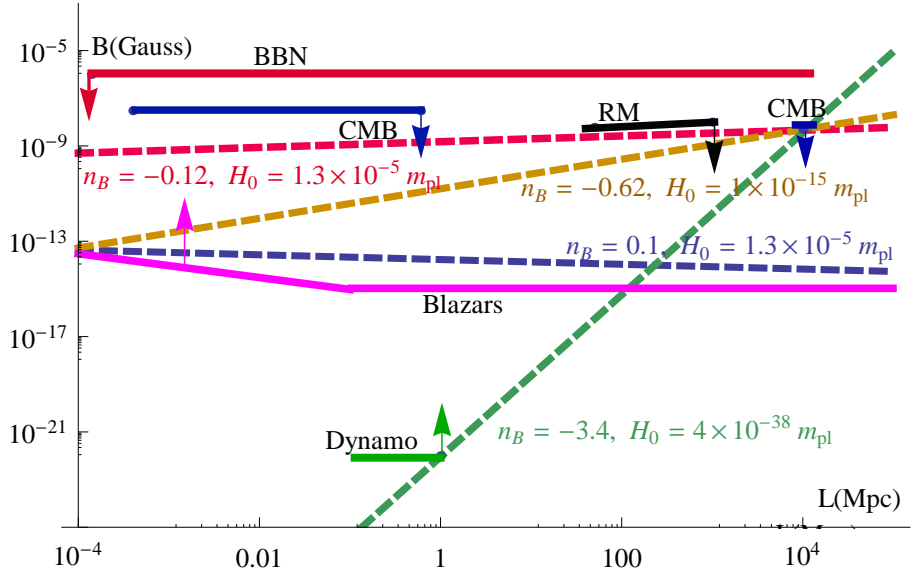}\\
\caption{Many admissible spectrums of magnetic fields. If we take into account the more
restrictive constrain that comes from TeV Blazars we obtain that our spectrums should
remain close to scale invariance. The most blue tilted compatible spectrum is for
$n_B\simeq 0.1$ and a scale $H_0=1.3\times 10^{-5}m_{pl}$. Furthermore the most red
tilted magnetic fields are for $n_B\simeq -0.62$ and a lower scale $\sim
10^{-15}m_{pl}$ }
  \label{FigB(varios)}
  \end{figure}

\end{document}